\documentclass[12pt]{article}

\usepackage{graphicx, float, array, xspace, amscd, amsmath, amsthm, amssymb, amsfonts, latexsym, relsize, bbm}

\usepackage{setspace}
\usepackage[letterpaper, margin=1in]{geometry}
\numberwithin{equation}{section}

\usepackage[bbgreekl]{mathbbol}
\usepackage[T1]{fontenc}
\usepackage[sort&compress, comma, square, numbers]{natbib}
\usepackage{tensor}

\usepackage[linktocpage=true]{hyperref}
\hypersetup{colorlinks=true, citecolor=blue, linkcolor=blue, urlcolor=blue, pdfauthor={}, pdftitle={},breaklinks=true}

\renewcommand{\d}{\delta}
\newcommand{\e}{\epsilon}
\newcommand{\del}{{\partial}}
\newcommand{\delb}{{\bar\partial}}
\newcommand{\End}{{\text{End}\,}}
\newcommand{\dd}{{\text{d}}}
\DeclareMathOperator{\tr}{Tr}

\begin{document}
\pagestyle{empty}      
\begin{center}
\null\vskip0.2in
{\Huge  A Heterotic K\"ahler Gravity\\ and the Distance Conjecture\\[0.5in]}
{Javier Jos\'{e} Murgas Ibarra$^{a}$, Paul-Konstantin Oehlmann$^{b}$, Fabian Ruehle$^{b,c,d}$, \\
Eirik Eik Svanes$^{a}$\\[0.5in]}
{\it 
$^a$Department of Mathematics and Physics \\
Faculty of Science and Technology, University of Stavanger\\
N-4036, Stavanger, Norway\\[3ex]
$^b$ Department of Physics, Northeastern University,\\
Boston, MA 02115, USA\\[3ex]
$^c$ Department of Mathematics, Northeastern University,\\
Boston, MA 02115, USA\\[3ex]
$^d$ NSF Institute for Artificial Intelligence and Fundamental Interactions,\\
Boston, USA
}
\end{center}
\vspace{1cm}
\begin{abstract}
Deformations of the heterotic superpotential give rise to a topological holomorphic theory with similarities to both Kodaira-Spencer gravity and holomorphic Chern-Simons theory. Although the action is cubic, it is only quadratic in the complex structure deformations (the Beltrami differential). Treated separately, for large fluxes, or alternatively at large distances in the background complex structure moduli space, these fields can be integrated out to obtain a new field theory in the remaining fields, which describe the complexified hermitian and gauge degrees of freedom. We investigate properties of this new holomorphic theory, and in particular connections to the swampland distance conjecture in the context of heterotic string theory. In the process, we define a new type of symplectic cohomology theory, where the background complex structure Beltrami differential plays the role of the symplectic form. 
\end{abstract}

\clearpage
\pagestyle{empty}
\setstretch{1.0}
\tableofcontents
\clearpage

\setcounter{page}{1}
\pagestyle{plain}
\setstretch{1.2}

\section{Introduction} 
Understanding the circumstances under which quantum field theories can be UV completed with gravity is one of the main efforts of the last decades of theoretical physics.
Mayor inspirations for consistent UV completions have been drawn from string theory, which is a UV complete theory of quantum gravity.

The swampland program~\cite{Vafa:2005ui} (see \cite{vanBeest:2021lhn,Grana:2021zvf,Agmon:2022thq} for recent reviews) aims to extract and connect necessary conditions for UV completion. The swampland distance conjecture~\cite{Ooguri:2006in} is a well-studied conjecture with a lot of evidence supporting it, stating that EFTs with a moduli space $\mathcal{M}$ get heavily modified when moving to the boundary of $\mathcal{M}$, which eventually leads to a breakdown of its description~\cite{Ooguri:2006in}. More concretely, it is conjectured that there is a tower of light states whose mass gap closes exponentially fast towards infinite geodesic distance in $\mathcal{M}$. 
These limits have been studied extensively in the recent literature and further sharpened over time (see \cite{Etheredge:2022opl} and reference therein),  mostly in theories with more than four supercharges, where BPS states give a lot of extra control. In such setups, the respective towers could be characterized, and it was further proposed that their origin should be attributed to either light Kaluza-Klein states or light string degrees of freedom. This is known as the emergent string conjecture~\cite{Lee:2019wij}.

Four-dimensional theories with $\mathcal{N}=1$ supersymmetry are more challenging, since there are fewer BPS-protected states, and the moduli spaces may receive non-perturbative corrections. Such cases were studied in F-theory and Type II orientifolds for asymptotic limits in the K\"ahler and complex structure moduli spaces of the underlying CY three- and four-folds. In contrast, infinite distance limits in heterotic string theory setups are less well understood and explored. Efforts are mostly focusing on the K\"ahler moduli space \cite{Brodie:2021ain,Brodie:2021nit,Lanza:2021udy,Castellano:2023jjt} and invoke mirror symmetry for analogous statements for complex structure moduli, similar to the Type II scenarios. 

The additional complication in the heterotic string stems from the fact that it requires choosing an $SU(3)$ structure manifold $X_3$ \textit{together with} a compatible vector bundle $V$. As a consequence, the moduli space of a heterotic string theory with a choice of $(X_3,V)$ contains bundle deformations of $V$ in addition to K\"ahler and complex structure deformations. Moreover, the defining data $X_3$ and $V$ and their moduli spaces are not independent quantities but constrained by the Bianchi identity and the hermitian Yang-Mills equations, or more generally the Hull-Strominger (HS) system \cite{Hull:1986kz, Strominger:1986uh}. This makes studying the full heterotic string moduli space and its various limits difficult, even at finite distances. Setting the gauge bundle $V$ equal to the tangent bundle $TX_{3}$ (the so called \textit{standard embedding}) solves the HS system trivially and leads to a local split of complex structure and K\"ahler moduli spaces, but is far from being the most generic situation. Moreover, due to the enhanced $\mathcal{N}=(2,2)$ supersymmetry on the world sheet (WS), this is also the only instance in which mirror symmetry for the heterotic string is well-understood and can be used to probe infinite distance limits. In the generic situation, WS supersymmetry is broken to $\mathcal{N}=(0,2)$ and the mirror map in this case is not known in general.\footnote{Progress on establishing a $(0,2)$ mirror map include earlier cases, when $V$ is a perturbation of the tangent bundle of $X_3$ \cite{Melnikov:2010sa,Melnikov:2012hk}, and more recent generalizations \cite{Gu:2019byn}.} Hence we need to study the heterotic moduli space on $(X_3,V)$ without invoking mirror symmetry.   

These complications make the heterotic moduli space intimidating at first glance. However, fairly recently a better understanding of the generic heterotic moduli problem has emerged~\cite{Anderson:2010mh, Anderson:2011ty, Anderson:2014xha, Garcia-Fernandez:2015hja, Ashmore:2018ybe, Ashmore:2019rkx, McOrist:2021dnd}. In particular, in \cite{Ashmore:2018ybe} it was shown that finite deformations of the system can be parameterized in terms of solutions to a Maurer-Cartan equation associated to a holomorphic Courant algebroid. This equation can be seen as the equation of motion of a quasi-topological theory derived from the heterotic superpotential. This theory has features resembling both Kodaira-Spencer gravity \cite{Bershadsky:1993cx} and Donaldson-Thomas theory, or holomorphic Chern-Simons theory \cite{donaldson1998gauge, thomas1997gauge}. In particular, it is a cubic theory in the fields, coupling the gauge degrees of freedom of holomorphic Chern-Simons theory to the gravitational modes of hermitian and complex structure degrees of freedom. 

Although the theory is cubic, it is only quadratic in the complex structure degrees of freedom, and a background flux gives rise to a mass term for these modes. At large fluxes, or correspondingly at large distance in the complex structure moduli space, these modes can be integrated out, leading to an effective theory in the remaining fields. In the presence of non-vanishing Yukawa couplings for the flux, when viewed as a closed $(2,1)$-form, and a superpotential term induced from holomorphic Courant algebroids (which are the natural mathematical structures that govern the symmetries of heterotic geometries \cite{Garcia-Fernandez:2015hja, Garcia-Fernandez:2018emx, Ashmore:2018ybe, Ashmore:2019rkx, Garcia-Fernandez:2020awc, Tellez-Dominguez:2023wwr, Streets:2024rfo, Silva:2024fvl}), we find a tower of light modes coming down at an exponential rate in geodesic distance as the complex structure modulus is taken to infinity, in accordance with the distance conjecture. 

These new effective topological heterotic K\"ahler gravity theories are also interesting from a mathematical point of view. In particular, exploring the gauge structure of these actions, and specifically their quadratic approximations, leads us to define a new type of ``symplectic cohomology'' theory, where the background flux/background complex structure plays the role of the symplectic form. We find that the gauge structure forms a natural part of new elliptic differential complexes, which can also be used to quantize the theories and study properties like anomalies and topological invariants.
 
This paper is structured as follows:
In Section~\ref{sec:reviewdistance}, we give a short introduction to the distance conjecture. We introduce the heterotic effective action that we study in this paper in Section~\ref{sec:EffTheory}. Section~\ref{sec:elliptic} is devoted to discussing important aspects of the action, such as ellipticity of the mass operator in  Sections~\ref{sec:4Dilaton1} - \ref{sec:4Dilaton3}. Inspired by the SYZ conjecture, we discuss in Section~\ref{sec:ToyModel} a toy model on a 3-sphere. In Section~\ref{sec:gauge}, we briefly look at a case that includes $\alpha'$ effects and thus gauge fields via the Chern-Simons term in the heterotic three-form field. We present our conclusions in Section~\ref{sec:conclusion}, and give more details about some of our calculations in three appendices.  

\section{The Distance Conjecture}
\label{sec:reviewdistance}
Within the swampland program, the distance conjecture \cite{Ooguri:2006in} is rather well-established and studied, in particular since it makes strong general predictions and connects to other conjectures, such as the absence of global symmetries or the weak gravity conjecture. The distance conjecture states that when we start with an EFT in quantum gravity at a certain point $P \in \mathcal{M}$ in its moduli space and move to another point $O \in \mathcal{M}$ that is at infinite distance along a shortest geodesic, $d(P,O)\to\infty$, there is an infinite tower of states that becomes exponentially light. The masses of this tower scale asymptotically as 
\begin{equation}
    m(O) \sim m(P) e^{-\alpha \cdot d(P,O)} \,,
\end{equation}
where $\alpha$ is a positive, order $\mathcal{O}(1)$ factor. The above behavior represents a critical breakdown of the EFT, which loses validity in this limit. One source for the tower of states are KK modes that becomes light when approaching a decompactification limit. A second source are light string states. According to the \textit{emergent string conjecture}~\cite{Lee:2019wij,Lee:2019xtm}, these are the only two possibilities. The prototypical example is the compactification of a $D$-dimensional theory on a circle. This theory has a one-dimensional moduli space $\mathcal{M}$ corresponding to the radius of the circle. The dimensional reduction of the radon field $R$ then has the kinetic term (in Einstein frame) 
\begin{equation}
    S= \int \left(\frac{d R}{R}\right)^2 + \ldots  \, .
\end{equation}
The field space metric is thus given by $1/R$ and distances between the points $P$ and $O$ with radii $R_P$ and $R_O$ are
\begin{align}
d(R_P,R_O)=\int_{R_P}^{R_O} \frac{1}{R} =\text{log}(R_O/R_P)\, ,
\end{align}
and hence $R_O \sim e^{d(R_P,R_O)}$. There is indeed already one tower we know of, which is the KK tower whose masses scale as 
\begin{align}
    m_n = \frac{n}{R} \sim n e^{-d(R_P,R_O)} \, ,
\end{align}
as expected. There does also exist a second limit, for  $R_O \rightarrow 0$ which is at infinite distance.
Field-theoretically, it is not clear what type of characteristic objects become light in this limit, but from string theory we know that the light states are the string winding modes, whose masses scale as  
\begin{align}
    m_{\hat{n}} =\frac{ \hat{n} R_O }{ \alpha^\prime} \,. 
\end{align}
These states become light at $R_O \rightarrow 0$, and correspond to the  T-dual description of the $R_O \rightarrow \infty $ limit.  

This is perhaps the simplest example possible: the moduli space is only one-dimensional and both types of towers can be identified straight-forwardly. Theories in lower dimensions and with less supersymmetry substantially enrich the discussion. For example, Calabi-Yau compactifications have $\mathcal{O}(100)$-dimensional complex structure and K\"ahler moduli spaces. The lower amount of supersymmetry makes it also more difficult to identify the infinite towers and their behaviors. 

\section{The Heterotic Effective Action}
\label{sec:EffTheory}
Throughout most of this paper we will set $\alpha'=0$, i.e., we will be working in the supergravity approximation where the gauge vector degrees of freedom decouple from the gravitational sector and the background geometry is Calabi-Yau. We will comment on the case $\alpha'\neq0$ in Section \ref{sec:gauge}, but leave a full treatment to future work. 

To set the stage, consider a Calabi-Yau manifold $(X,\Psi,g)$ with a fixed background metric $g$ and a holomorphic top-form $\Omega=e^{-2\phi}\Psi$, where $\phi$ denotes the dilaton. The heterotic superpotential reads \cite{LopesCardoso:2003dvb, Becker:2003gq, Gurrieri:2004dt}
\begin{equation}
    W=\int_X(H+i\dd\omega)\wedge\Omega\:,
\end{equation}
where $\omega$ is the Hermitian $(1,1)$-form which need not be closed off-shell. The closed three-form $H=\dd B$ is the heterotic NS-flux. Performing a finite but small fluctuation of this superpotential around a Calabi-Yau background, we get the following quasi-topological\footnote{The action is quasi-topological in that it does not depend on the background metric, but does depend on the background complex structure.} superpotential contribution \cite{Ashmore:2018ybe, Ashmore:2023vji}
\begin{equation}
    \Delta W= \int_X \left(x_a\bar\partial\mu^a+(H_{ab}+\partial x_{ab})\mu^a\mu^b+\mu^a\partial_ab\right)\wedge\Omega\:.
    \label{eq:SupAction}
\end{equation}
In this action, we have ignored off-shell deformations of the (axio-)dilaton. We will analyse the contribution from fluctuations of the dilaton in section \ref{sec:dilaton}.

The fields in the action \eqref{eq:SupAction} are as follows:
\begin{equation}
    x\in\Omega^{(0,1)}(T^{*(1,0)}X)
\end{equation}
is the complexified hermitian deformation, which also includes contributions from the Kalb-Ramond $B$-field.
\begin{equation}
\mu\in\Omega^{(0,1)}(T^{(1,0)}X)    
\end{equation}
is the complex structure deformation, or Beltrami differential. The field
\begin{equation}
    b\in\Omega^{(0,2)}
\end{equation}
derives from the holomorphic deformation of the $B$-field. Finally, by slight abuse of notation, 
\begin{equation}
    H\in\Omega^{(0,1)}(\wedge^2 T^{*(1,0)}X)=\Omega^{(2,1)}
\end{equation}
corresponds to the background flux, which is not a field but part of the background. The symmetries of the action requires $H$ to be both $\partial$-closed and $\bar\partial$-closed. This is in line with the holomorphic Courant algebroid picture of heterotic deformations, see \cite{Ashmore:2018ybe} for more details. Shifting $H$ by $\partial\bar\partial$-exact forms, we can, modulo a field redefinition of $x$, take $H$ to be Bott-Chern harmonic, which on a Calabi-Yau is the same as being harmonic ($H\in{\cal H}^{(2,1)}$) in the usual sense \cite{schweitzer2007autour}, and we will take $H$ harmonic for the rest of this paper.

In \eqref{eq:SupAction}, $H$ is treated as a background flux, which is quantized and hence cannot be deformed continuously. To connect our computations with the swampland distance conjecture, we need to take $H\in{\cal H}^{(2,1)}$ to be a direction in the background complex structure moduli space. Such a term is induced from a $(3,0)$-part of the three-form flux, which shifts the action~\eqref{eq:SupAction} to
\begin{equation}
    \label{eq:ShiftDeltaW}
    \Delta W \to \Delta W+c\int_X\frac{1}{3!} \mu^a\mu^b\mu^c\Omega_{abc}\wedge\Omega=\Delta W+c\int_X\Omega(\mu,\mu,\mu)\wedge\Omega\:.
\end{equation}
It is natural to include such a term both from a mathematical and phsyical point of view. Mathematically, it forms part of the more general holomorphic Courant algebroids, which give a natural mathematical description of six-dimensional heterotic geometries \cite{Garcia-Fernandez:2015hja, Garcia-Fernandez:2018emx, Ashmore:2018ybe, Ashmore:2019rkx, Garcia-Fernandez:2020awc, Tellez-Dominguez:2023wwr, Streets:2024rfo, Silva:2024fvl}. Physically, it can arise from Fermi condensates in heterotic supergravity~\cite{Dine:1985rz, derendinger1985low, LopesCardoso:2003sp, Manousselis:2005xa, Chatzistavrakidis:2012qb, Minasian:2017eur}, as explained in Appendix \ref{app:Condensates}. We will hence add this term in our study of the distance conjecture.

Turning off the background flux, the deformed superpotential becomes
\begin{equation}
    \Delta W= \int_X \left(x_a\bar\partial\mu^a+\partial x_{ab}\mu^a\mu^b+\mu^a\partial_ab+c\,\Omega(\mu,\mu,\mu)\right)\wedge\Omega\:,
    \label{eq:SupAction2}
\end{equation}
where we have included the term cubic in complex structure deformations. Below, we will treat this as a quantum theory in the complex structure deformations. Let us therefore expand $\mu$ around a background on-shell complex structure, and set
\begin{equation}
    \mu\rightarrow\mu_0+\mu\:,
\end{equation}
where $\mu_0$ satisfies the Maurer-Cartan equation
\begin{equation}
    \bar\partial\mu^a_0+\tfrac12[\mu_0,\mu_0]^a=0\:.
    \label{eq:MCeq}
\end{equation}
It is well-known that solutions to \eqref{eq:MCeq} and hence the local complex structure moduli space are parametrized by $H^{(2,1)}\cong {\cal H}^{(0,1)}(T^{(1,0)}(X))$. In particular, in a parametrization in which $\mu_0$ is {\it harmonic}, both terms in \eqref{eq:MCeq} vanish separately. To see this, note that by contracting \eqref{eq:MCeq} with $\Omega$, the second term becomes
\begin{equation}
    \tfrac12[\mu_0,\mu_0]^a\Omega_{abc}\dd z^{bc}\propto \partial\left(\mu_0^a\mu_0^b\Omega_{abc}\dd z^c\right)=0\:.
\end{equation}
The last equality follows from special geometry, and the fact that $\mu_0^a\mu_0^b\Omega_{abc}\dd z^c$ is harmonic~\cite{Candelas:1990pi, Strominger:1990pd}.

Expanding \eqref{eq:SupAction2} in quantum fluctuations $\mu$ around the background $\mu_0$, we then get
\begin{equation}
    \Delta W= \int_X \left(x_a\bar\partial\mu^a+\partial x_{ab}\mu^a\mu^b+\mu^a\partial_ab+3c\,\Omega(\mu_0,\mu,\mu)\right)\wedge\Omega\:+{\cal O}(\mu^3)\,,
    \label{eq:SupAction4}
\end{equation} 
up to order $\mu^2$. We dropped terms linear in the fields, as these only serve to correct the equations of motion of the background field, as well as a constant contribution from $\Omega(\mu_0,\mu_0,\mu_0)$. The latter becomes relevant when we consider the dilaton coupling in the next sections.

Note that the last term in \eqref{eq:SupAction4} is exactly of the same form as the $H$-term of \eqref{eq:SupAction}, so we will identify them from hereon. That is, upon an appropriate redefinition of $\mu_0$ absorbing $c$, this term becomes precisely the $H$-term of \eqref{eq:SupAction}. This is what we mean when we say we treat $H$ as a background complex structure modulus.

\subsection{The Effective Action}
\label{sec:DiscussEFF}
Let us treat the superpotential contribution $\Delta W$ of \eqref{eq:SupAction} as a quantum theory and study its implications. For example, note that the expression is linear in $x$, so integrating out $x$ simply imposes its equation of motion which puts the complex structure deformations on-shell. That is, they satisfy the Kodaira--Spencer Maurer--Cartan equation \eqref{eq:MCeq}. More interestingly, one can ask what happens when we integrate out $\mu$? As the action is quadratic in $\mu$, there is hope of performing such an integral explicitly, leading to an effective theory for the hermitian degrees of freedom $x$ and $b$. In a usual field theory, the quadratic term is given in terms of an elliptic differential operator with a finite-dimensional kernel. Hence, the operator is invertible, modulo this kernel. Our action \eqref{eq:SupAction} does not have a kinetic term for~$\mu$. However, if $H$ is invertible when acting point-wise as a matrix operator on $\mu$ valued in $\Omega^{(0,1)}(T^{(1,0)})$, we can nevertheless employ the fundamental theorem of quantum field theory. Indeed, this quadratic term is a mass term for $\mu$, and for large values of $H$ the fluctuations $\mu$ can be integrated out. We shall see below that $H$ is invertible if and only if $H$ has non-vanishing Yukawa triple coupling, and so we make this requirement when we integrate out~$\mu$.

When integrating out $\mu$, we treat it as a dynamical field in the path integral, with $\Delta W$ as the action 
\begin{align}
Z=e^{-S_\text{eff}}=\int \mathcal{D} \mu\, e^{-\Delta W } \, ,
\end{align}
which then gives rise to the effective action $S_\text{eff}$. Integrating out $\mu$ is then a standard exercise in field theory. Some straight-forward manipulations show that the action \eqref{eq:SupAction} can be written as
\begin{equation}
\label{eq:CompAction}
    \Delta W= i\int_X\left(\tfrac12(\delb x+\del b)_{a\bar c\bar d}\Omega^{\bar c\bar d\bar e}\mu^a_{\bar e} + \mu^{a}_{\bar d}(H+\del x)_{ab\bar c}\Omega^{\bar c\bar d\bar e}\mu_{\bar e}^b\right){\rm dvol}\:,
\end{equation}
where the volume form is
\begin{equation}
    {\rm dvol}=\frac{i}{\vert\Omega\vert^2}\Omega\wedge\bar\Omega\:.
\end{equation}
Let us define a symmetric pairing $H+\del x$ on $\Omega^{(0,1)}(T^{(1,0)}X)$ as
\begin{equation}
\label{eq:SymPair1}
    {(H+\del x)_a}^{\bar d}{\,_b}^{\bar e}=\tfrac{2i}{\vert\Omega\vert^2}(H+\del x)_{ab\bar c}\Omega^{\bar c\bar d\bar e}\:,
\end{equation}
where we have chosen a convenient normalization, where the norm appears in the denominator so as to keep this matrix topological, i.e., metric independent. Note that the norm $\vert\Omega\vert^2$ is constant on a Calabi-Yau. As invertibility is an open property, for small enough $x$ this matrix is invertible provided $H$ is invertible when viewed as a symmetric pairing on $\Omega^{(0,1)}(T^{(1,0)}X)$.

We also define an element $(\delb x+\del b)\in\Omega^{(1,0)}(T^{(0,1)}X)$ as
\begin{equation}
    (\delb x+\del b)^{\bar e}_a=\tfrac{i}{2\vert\Omega\vert^2}(\delb x+\del b)_{a\bar c\bar d}\Omega^{\bar c\bar d\bar e}\:,
\end{equation}
with a convenient normalization so that this expression also remains topological. The action~\eqref{eq:CompAction} may then be written in a more conventional, and and manifestly topological form as
\begin{equation}
    \Delta W= \int_Xi\Omega\wedge\bar\Omega\left((\delb x+\del b)\cdot\mu + \tfrac12\mu\cdot (H+\del x)\cdot\mu\right)\:,
\end{equation}
where we have suppressed the indices. To further simplify the notation, we also drop the topological measure $i\Omega\wedge\bar\Omega$, and simply write the topological integration as
\begin{equation}
\label{eq:TopMeasure}
    \int_X i\Omega\wedge\bar\Omega\;\rightarrow\;\int_X\:.
\end{equation}
If not specified otherwise, this is the measure we use when we integrate functions over $X$. Integrating top-forms is done in the normal way. 

We can now solve the path integral by first completing the square, and letting
\begin{align}
    \mu\; \rightarrow \;\mu - (H+\del x)^{-1}\cdot (\delb x+\del b)\:.
\end{align}
This shift eliminates the linear term in $\mu$ such that we can solve the Gauss integral. Note also that such a linear shift will not change the path integral measure. We then obtain the new action
\begin{align}
    \Delta W= \int_X\left(\mu\cdot (H+\del x)\cdot\mu -\tfrac12 (\delb x+\del b)\cdot (H+\del x)^{-1}\cdot (\delb x+\del b)\right)\:. 
\end{align}
After this shift, the action is purely quadratic in $\mu$ and we can therefore integrate out the fluctuations $\mu$. This results in an effective field theory for the remaining degrees of freedom of the form  
\begin{align}
    S_\text{eff} &= \tfrac12\log\left(\det \left(H+\partial x\right)\right)-\tfrac12\int_X ( \bar\partial x +\partial b)\cdot (H+\partial x)^{-1}\cdot (\bar\partial x +\partial b)\notag\\
    &=\tfrac12\int_X\tr\left(\log(H+\partial x)\right)- \tfrac12\int_X ( \bar\partial x +\partial b)\cdot (H+\partial x)^{-1} \cdot(\bar\partial x +\partial b)\:,
\label{eq:ExactEffAction}
\end{align}
where the trace in the first term is over matrix indices of the matrix $(H+\del x)$, as defined in~\eqref{eq:SymPair1}. 

Note that this is an exact theory for the hermitian mode $x$ and the $b$-field. By naive inspection, the action has the following gauge symmetries
\begin{align}
    \delta x&= \partial\alpha \label{gau1}\\
    \delta b&= \bar\partial\alpha\:, \label{gau2}
\end{align}
where $\alpha\in\Omega^{(0,1)}(X)$. However, as we discuss in Section \ref{sec:elliptic} this is not quite accurate. Note also that since our starting theory is topological (it does not depend explicitly on the background metric), this new effective theory is topological as well.\footnote{As noted above, we have taken $H$ to be harmonic, which does imply an implicit dependence on the background metric. It should however be noted that, upon an appropriate field redefinition of $x$, $H$ in \eqref{eq:ExactEffAction} may be shifted  by $\partial\bar\partial$-exact forms to any other representative of the Bott-Chern class of $H$. The Bott-Chern class is quasi-topological, i.e., it only needs a complex structure to be well-defined~\cite{schweitzer2007autour}.}

Expanding the theory~\eqref{eq:ExactEffAction} to quadratic order in the fields $x$ and $b$, we get
\begin{align}
           S_\text{eff} &=\tfrac12\int_X\tr\Big(\log\left(H\right)+\tfrac12\left(H^{-1}\partial x\right)-\tfrac14\left((H^{-1}\partial x)^2\right)\Big)\notag\\
            &-\tfrac12\int_X ( \bar\partial x +\partial b)\cdot (H)^{-1} \cdot (\bar\partial x +\partial b)+ \mathcal{O}(x^3,b^3)\,.
\end{align}
We now study the invertibility of $H$. Note first that we can also view $H$ as a harmonic element $h\in{\cal H}^{(0,1)}(T^{(1,0)}X)$ via
\begin{equation}
    H=\tfrac12H_{ab\bar c}dz^{ab\bar c}=\Omega(h)=\tfrac12 h^c\wedge\Omega_{cab} dz^{ab}\:.
\end{equation}
Since $h_{\bar a}^{d}$ is harmonic, $g^{a\bar a}h_{\bar a}^{d}=h^{ad}$ is symmetric \cite{Candelas:1990pi}. This quantity can then serve as a ``metric'', i.e., an object with which can be used to raise and lower indices, though we will not use this feature much in the current paper. It follows given \eqref{eq:SymPair1} that 
\begin{equation}
    {H_a}^{\bar d}{\,_b}^{\bar e}=\tfrac{2i}{\vert\Omega\vert^2}H_{ab\bar c}\Omega^{\bar c\bar d\bar e}=\tfrac{2i}{\vert\Omega\vert^2}h^f_{\bar c}\Omega_{fab}\Omega^{\bar c\bar d\bar e}\:.
\end{equation}
The ``determinant'' $\vert h\vert$ of $h_{\bar a}^b$ is given by
\begin{equation}
\label{eq:deth}
    \vert h\vert\bar\Omega=
    \frac{1}{3!}h^b\wedge h^d\wedge h^f\;\Omega_{bdf}=\Omega(h,h,h)\,.
\end{equation}
Note that $\vert h\vert$ is also proportional to the determinant of the symmetric matrix $h^{ab}$, where the proportionality factor depends on the determinant of the Calabi-Yau metric. Equation~\eqref{eq:deth} is then the Yukawa coupling of the harmonic form $h$, which is known to be a constant times~$\bar\Omega$ \cite{Candelas:1990pi, Strominger:1990pd}. Therefore, if $h$ has a non-zero Yukawa coupling, the determinant $\vert h\vert$ becomes a point-wise non-zero constant. That is, $h_{\bar a}^b$ is invertible whenever this constant (proportional to the Yukawa coupling of $h$) is non-zero, as mentioned above. Let then $h_a^{\bar b}$ denote its inverse,
\begin{equation}
    h_{\bar a}^{c}h_c^{\bar b}=\delta^{\bar a}_{\bar b}\:.
\end{equation}
As we show in Appendix \ref{app:Geometry}, $h^{-1}$ is harmonic if $h$ is. We also introduce a rescaled $\Omega$-symbol
\begin{equation*}
    \tilde\Omega_{abc}={\vert h\vert}^{-\tfrac12}\Omega_{abc}\:,
\end{equation*}
which satisfies the identity
\begin{equation}
    h_{\bar a}^{d}\;\tilde\Omega^{\bar a\bar b\bar c}\;\tilde\Omega_{def}
    =\vert\Omega\vert^2\left(h_e^{\bar b}h_f^{\bar c}-h_f^{\bar b}h_e^{\bar c}\right)\:,
\end{equation}
where
\begin{equation}
    \vert\Omega\vert^2=\tfrac{1}{3!}\Omega^{\bar a\bar b\bar c}\bar\Omega_{\bar a\bar b\bar c}\:.
\end{equation}
Using this, it is straightforward to find the inverse of $H$ as
\begin{equation}
    \label{eq:Hinv}
    {(H^{-1})_{\bar a}}^b{{\,}_{\bar c}}^d=\tfrac{1}{2i\vert h\vert}\left(\tfrac12h_{\bar a}^b h_{\bar c}^d-h_{\bar c}^b h_{\bar a}^d\right)\:.
\end{equation}
Note that also $H^{-1}$ is topological, in the sense that there is no explicit metric dependence. 

The quadratic part of the new effective action then becomes
\begin{equation}
\label{eq:effect}
    S_\text{eff} = \tfrac12\int_X\tr\left(H^{-1}\partial x\right)-\tfrac14\int_X\tr\left((H^{-1}\partial x)^2\right)-\tfrac12\int_X ( \bar\partial x +\partial b)_b^{\bar a} {(H^{-1})_{\bar a}}^b{{\,}_{\bar c}}^d (\bar\partial x +\partial b)_d^{\bar c}\:,
\end{equation}
up to the constant term independent of $x$. Recall that
\begin{equation}
    (\bar\partial x +\partial b)_a^{\bar d}
    =\tfrac{i}{2\vert\Omega\vert^2}\left(\delb x + \del b\right)_{a\bar b\bar c}{\Omega}^{\bar b\bar c\bar d}\:.
\end{equation}
The first term linear in $x$ in the action~\eqref{eq:effect} actually drops out: As one can show using \eqref{eq:Hinv}, it is proportional to the inner product of $\partial x$ with $\overline{\Omega(h,h)}$, where
\begin{equation}
    \Omega(h,h)=\tfrac12\Omega_{abc}h^a\wedge h^b\wedge\dd z^c\:.
\end{equation}
Special geometry dictates that $\Omega(h,h)$ is harmonic \cite{Candelas:1990pi, Strominger:1990pd}, so this inner product vanishes. 

The quadratic effective action therefore becomes
\begin{equation}
\label{eq:effect2}
    S_\text{eff} = -\tfrac14\int_X\tr\left((H^{-1}\partial x)^2\right)-\tfrac12\int_X ( \bar\partial x +\partial b)_b^{\bar a} (H^{-1}){}_{\bar a}{}^b{}_{\bar c}{}^d (\bar\partial x +\partial b)_d^{\bar c}\:,    
\end{equation}
The first and second term in the effective action are of order ${\cal O}(H^{-2})$ and ${\cal O}(H^{-1})$, respectively. Treating $S_\text{eff}$ as an effective superpotential, we therefore expect two towers of light states to come down at different rates as we send $H\rightarrow\infty$. We will analyze this in further detail below.

\subsection{Heterotic action including the dilaton}
\label{sec:dilaton}
In the above we have neglected complications due to fluctuations of the dilaton. In string theory, the dilaton is a field and we need to take its fluctuations into account. Deforming the superpotential including the dilaton results in the action
\begin{equation}
\label{eq:SDilaton}
S= \int_X(dx+H+db) \wedge e^\chi(\Omega + \Omega(\mu)+\Omega(\mu,\mu))+c\int_Xe^\chi\Omega(h,h,h)\wedge\Omega\:,
\end{equation}
where $\chi$ denotes the complexified deformation of the dilaton, or the axio-dilaton. Note that we have now included the ``constant'' background term proportional to $\Omega(h,h,h)$, as this term also couples to the axio-dilaton. Using the Hodge decomposition of $\chi$, we can take $\chi$ to be a total divergence, since any constant or harmonic part can be absorbed by rescaling the action. Using equation \eqref{eq:deth}, we then obtain the action
\begin{align}
    S= \int_X \left(e^{\chi}( \bar{\partial}x_a+{\partial_a}b)\mu^a +e^{\chi}(H_{ab}+\partial x_{ab})\mu^a \mu^b + e^\chi\bar{\partial}b +e^\chi c\,\vert h\vert\bar\Omega \right) \wedge \Omega\:.
    \label{eq:SwithDilaton}
\end{align}
We next integrate out $\mu$ in the quadratic action following the same procedure as above, which results in
\begin{align}
     S_\text{eff} &=\int_X\tfrac12\tr\log\left[e^{\chi} \left(H+\partial x\right)\right]\notag \\
     &-\tfrac12\int_Xe^\chi ( \bar\partial x +\partial b)\cdot (H+\partial x)^{-1}\cdot (\bar\partial x +\partial b)+\int_Xe^\chi\left(\bar{\partial}b+c\,\vert h\vert\bar\Omega \right) \wedge \Omega\:.
     \label{eq:effect3}
\end{align}
Note that the axio-dilaton dependence of the first term actually vanishes, since 
\begin{equation}
    \label{eq:TrChiZero}
    \int_X\chi=0\:.
\end{equation}
This is a consequence of $\chi$ being a total divergence. The effective quadratic action can then be written as
\begin{equation}
    S_{\rm eff}=-\tfrac14\int_X\tr\left((H^{-1}\partial x)^2\right)-\tfrac12\int_X ( \bar\partial x +\partial b)\cdot (H)^{-1} \cdot(\bar\partial x +\partial b)+\int_X\left(\chi\bar{\partial}b+\tfrac{c}{2}\,\chi^2\,\vert h\vert\bar\Omega \right) \wedge \Omega\:,
\end{equation}
where again the linear term in the axio-dilaton multiplying $c\vert h\vert\bar\Omega$ drops out since the axio-dilaton is a total derivative. 

The last term in the action can be written as
\begin{equation}
    \int_X\left(\chi\bar{\partial}b+\tfrac{c}{2}\,\chi^2\,\vert h\vert\bar\Omega \right) \wedge \Omega=\int_X\left(i\,\chi{\cal L}_0(b)+\tfrac{i\,c\vert h\vert}{2}\chi^2\right)\:,
\end{equation}
where we are using the topological integration measure \eqref{eq:TopMeasure} and  
\begin{equation}
    {\cal L}_0(b)=\tfrac{i}{\vert\Omega\vert^2}(\Omega\lrcorner\delb b)=\tfrac{i}{\vert\Omega\vert^2}\left(\tfrac{1}{3!}\Omega^{\bar a \bar b\bar c}(\delb b)_{\bar a \bar b\bar c}\right)\:.
\end{equation}
Note that the $\chi^2$-term in the action, proportional to the Yukawa coupling, gives rise to a large mass-term for the axio-dilaton as $\vert h\vert\rightarrow\infty$. Just like for $\mu$, this also freezes out higher order momentum modes of the dilaton, and we can effectively integrate it out from the quadratic action. This then results in the equation of motion
\begin{equation}
    \label{eq:EoMBField1}
    {\cal L}_0(b)+i\,c\vert h\vert\chi=0\:.
\end{equation}
Solving this for the dilaton, and inserting the result in the action gives
\begin{equation}
\label{eq:Fulleffect}
    S_{\rm eff}=-\int_X\tfrac14\tr\left((H^{-1}\partial x)^2\right)-\tfrac12\int_X ( \bar\partial x +\partial b) (H)^{-1} (\bar\partial x +\partial b)+\tfrac{i}{2c\vert h\vert}\int_X{\cal L}_0(b)^2\:.
\end{equation}
Note that the first term goes as ${\cal O}(H^{-2})$, the second term goes as ${\cal O}(H^{-1})$, and the third term goes as ${\cal O}(H^{-3})$. We expect a tower of light modes coming down at each of these respective rates. Comparing with the previous two towers, we see that including the axio-dilaton produced an additional tower of states that becomes light at rate of ${\cal O}(H^{-3})$. 

The action \eqref{eq:Fulleffect} is difficult to analyze, especially the middle term coupling the hermitian modes given by $x$ with the $b$-field. The term simplifies in backgrounds where $b=0$, or where the dilaton mass scale is far above the other fields, such that we expand the dilaton to linear instead of quadratic order. In this case, the dilaton equation of motion becomes
\begin{equation}
    \bar\partial b=0\:,
\end{equation}
which on a Calabi-Yau background means that the $(0,2)$-form $b$ is $\bar\partial$-exact. After an appropriate field redefinition that shifts $x$ by a $\partial$-exact form, the quadratic action becomes
\begin{align}
    S_\text{eff} &= -\tfrac14\tr\left((H^{-1}\partial x)^2\right)-\tfrac12\int_X ( \bar\partial x)_b^{\bar a} (H^{-1}){}_{\bar a}{}^b{}_{\bar c}{}^d (\bar\partial x)_d^{\bar c}\notag\\
    &=\int_X \del x \star_H \del x + \delb x \star'_{H} \delb x \:,
\label{eq:effect4}
\end{align}
where the last equality serves to define ``Hodge duals'' $\star_H$ and $\star'_H$ that are constructed using $H$ instead of $g$. Note that these Hodge duals give rise to symmetric pairings. Explicitly, the Hodge duals take the form
\begin{align}
    \star_H \del x&\quad\propto\quad{(H^{-1})_{\bar a}}^b{{\,}_{\bar c}}^d{(H^{-1})_{\bar f}}^e{{\,}_{\bar h}}^j\Omega^{\bar l \bar c\bar f}\Omega_{bjk}\,\del x_{de\bar l}\,\dd z^{k\bar a\bar h}\:,\\
    \star'_{H} \delb x&\quad\propto\quad(H^{-1}){}_{\bar a}{}^b{}_{\bar c}{}^d\Omega_{bef} (\bar\partial x)_d^{\bar c}\,\dd z^{ef\bar a}\:,
\end{align}
where we suppressed unimportant numerical prefactors.

The action \eqref{eq:effect4} is slightly different from \eqref{eq:effect2}. In particular, the new action appears to have a naive gauge symmetry $\delta x=\partial\bar\partial f$ for some function $f$, though this again turns out to be not quite accurate, which we discuss in Section~\ref{sec:elliptic}.

We observe that the quadratic action \eqref{eq:effect4} has again two different terms of order $\mathcal{O}(H^{-2})$ and $\mathcal{O}(H^{-1})$, respectively. As we send the background field $H$ to infinity and approach the boundary of complex structure moduli space, we thus expect two towers of massive states coming down at different rates. In particular, the kinetic operator of the theory takes the form
\begin{equation}
    \Delta_H=\del\star_H \del+\delb \star'_{H} \delb\:.
\end{equation}
Note that this operator has no explicit dependence on the choice of background metric, reflecting the topological nature of the theory. However, this operator maps $(1,1)$-forms to $(2,2)$-forms. In order to discuss its eigenvalues, we need to map back to $(1,1)$-forms. This is perhaps most conveniently done by choosing a fixed background metric, taking the ordinary Hodge-dual $*$, and considering the operator
\begin{equation}
    *\Delta_H:\;\;\Omega^{(1,1)}\rightarrow\Omega^{(1,1)}\:.
\end{equation}
Note that the numerical values of the individual eigenvalues depend on the choice of fixed metric, but their overall behavior as we approach a point at large distance in complex structure moduli space is universal. 

However, there is another slight complication, stemming from modes that arise from a combination of both terms in the operator. Since the two operators in $\Delta_H$ do not commute, they cannot be diagonalized simultaneously. However, the corresponding corrections to the eigenvalues will be of higher order in the $H^{-1}$ expansion, so that we still get two separate towers of modes. More concretely, by picking a K\"ahler metric on $X$, we can Hodge-decompose $x$ as
\begin{equation}
    x=x_h+\del\delb f+\del^\dagger\delb\alpha+\delb^\dagger\left(\del\beta+\del^\dagger\gamma\right)\:,
\end{equation}
where $x_h$ is harmonic. The last term appears in both parts of the action \eqref{eq:effect4}, while the second term only appears in the first part of the action, and it is this term that gives the leading contribution to eigenvalues of $\mathcal{O}(H^{-2})$. To illustrate why, consider a family of $2\times 2$ matrices 
\begin{equation}
    A_t=\begin{pmatrix}
ta & tb\\
tc & d
\end{pmatrix}\:,
\end{equation}
where $ta$ models the part of $\Delta_H$ acting on the space of forms of the type $\del^\dagger\delb\alpha$, while $d$ models the part of $\Delta_H$ acting on the space of forms of the type $\delb^\dagger\left(\del\beta+\del^\dagger\gamma\right)$. Here, $t$ is a small parameter modeling the separation of scales. It is straight-forward to compute the eigenvalues of $A_t$,
\begin{align}
    \lambda_1&=d+\mathcal{O}(t)\\
    \lambda_2&=ta+\mathcal{O}(t^2)\:,
\end{align}
which illustrates the separation of scales of eigenvalues. 

If we restrict ourselves for example to the tower of the heaviest modes, we can focus on the second part of \eqref{eq:effect4},
\begin{equation}
 \int_X \delb x \star'_{H} \delb x \:. 
\end{equation}
where, as noted above, $\star'_H$ is a Hodge-star like operator giving rise to a symmetric pairing. Classically, we may assume without loss of generality that $x=\delb^\dagger\gamma$ for a $\gamma\in\delb\Omega^{(1,1)}$. Using this, the action becomes
\begin{equation}
    \label{eq:SeffDilaton2}
    S_\text{eff} = \int_X \delb x \star'_H \delb x=\int_X\gamma\Delta\star'_H\Delta\gamma\:,
\end{equation}
where $\Delta$ is the Laplacian. Since $\Delta$ is elliptic and $\star'_H$ is invertible, the operator $\Delta\star'_H\Delta$ is elliptic. It will hence have a finite-dimensional kernel, and a discrete tower of eigenmodes with a non-zero eigenvalue that tends to zero as $H\to\infty$ at a rate of $\mathcal{O}(H^{-1})$. This further translates to an exponential decay in geodesic distance using the moduli space metric as we discuss in Section~\ref{sec:infdistance}.

\subsection{Integrating out the dilaton from the exact action}
\label{sec:DilatonInt}
In Section~\ref{sec:dilaton}, we implicitly assumed that all our fields live at roughly the same energy scale, and thus we expand all the fields to at most quadratic order.  If the natural energy scale of the dilaton is far below the other fields, we also need to consider higher momentum modes of the dilaton, and hence of $\chi$. We should then integrate out the axio-dilaton from the exact action \eqref{eq:effect3}, instead of the action at quadratic order. Since the second and third terms in the effective action \eqref{eq:effect3} are linear in $e^{\chi}$, there is hope to integrate out the exponentiated dilaton from the action.

In integrating out the dilaton, we need to be careful with the Jacobian factor introduced in the path integral measure when changing variables from $\chi$ to $e^{\chi}$. In position space, where $p\in X$ is a point in $X$, the path integral over the dilaton is
\begin{equation}
    \int{\cal D}\chi=\prod_p\int\dd\chi_p=\prod_p\int\dd\left(e^{\chi_p}\right)e^{-\chi_p}=\int{\cal D}\left(e^{\chi}\right)e^{-\int_X\chi}=\int{\cal D}\left(e^{\chi}\right)\:.
\end{equation}
The last equality follows since the dilaton is taken to be a total derivative. 

The factor $e^{\chi}$ acts as a Lagrange multiplier in the action, which can be integrated out. We arrange the action into the form
\begin{align}
     S_\text{eff} &=\tfrac12\int_X\tr\log\left[\left(H+\partial x\right)\right]\notag \\
     &-\tfrac12\int_Xe^\chi ( \bar\partial x +\partial b) \cdot(H+\partial x)^{-1}\cdot (\bar\partial x +\partial b)+\int_Xe^\chi\left(\bar{\partial}b+c\,\vert h\vert\bar\Omega \right) \wedge \Omega\notag\\
     &=\tfrac12\tr\log\left[\left(H+\partial x\right)\right]+\int_Xe^\chi{\cal L}(x,b)\notag\\
     &=\tfrac12\tr\log\left[\left(H+\partial x\right)\right]+\int_Xe^\chi\left({\cal L}_0(x,b)+i\,c\,\vert h\vert\right)\:,
\end{align}
where the equations are taken to define ${\cal L}_0(x,b)$ and ${\cal L}(x,b)$. Imposing the dilaton equation of motion, we get
\begin{equation}
    \left[e^\chi{\cal L}(x,b)\right]_{\textrm{total derivative}}=0\:.
\end{equation}
On a compact space, this is equivalent to $e^\chi{\cal L}(x,b)$ being constant,
\begin{equation}
    \dd\left(e^\chi{\cal L}(x,b)\right)=0\:,
\end{equation}
which gives
\begin{equation}
    (\dd\chi){\cal L}(x,b)+\dd {\cal L}(x,b) =0\:.
\end{equation}
We can solve this equation for $\chi$ to get
\begin{equation}
    \chi=-\left[\log({\cal L}(x,b))\right]_{\textrm{total derivative}}=-\log({\cal L}(x,b))+\left[\log({\cal L}(x,b))\right]_{\textrm{constant}}\:,
\end{equation}
where the constant part of $\log({\cal L}(x,b))$ is given by
\begin{equation}
    \left[\log({\cal L}(x,b))\right]_{\textrm{constant}}=\frac{1}{\vert\Omega\vert^2{\rm Vol}(X)}\int_X\log({\cal L}(x,b))\:,
\end{equation}
where the norm of $\Omega$ comes from the topological measure \eqref{eq:TopMeasure} in the integration. Inserting this back into the action gives
\begin{equation}
    \int_X e^\chi{\cal L}(x,b)=\vert\Omega\vert^2{\rm Vol}(X)\exp\left(\frac{1}{\vert\Omega\vert^2{\rm Vol}(X)}\int_X\log({\cal L}(x,b))\right)\:.
\end{equation}
Expanding expand the logarithm as
\begin{equation}
    \log({\cal L}(x,b))=\log({\cal L}_0(x,b)+i\,c\,\vert h\vert)=\log(i\,c\,\vert h\vert)+\tfrac{1}{i\,c\,\vert h\vert}{\cal L}_0(x,b)+\tfrac{1}{2c^2\,\vert h\vert^2}{\cal L}_0(x,b)^2+\ldots\:,
\end{equation}
the action becomes
\begin{align}
    \int_X e^\chi{\cal L}(x,b)=i\,c\,\vert h\vert\vert\Omega\vert^2{\rm Vol}(X)\exp\Bigg(\frac{1}{\vert\Omega\vert^2{\rm Vol}(X)}\int_X&\Big[\tfrac{1}{i\,c\,\vert h\vert}{\cal L}_0(x,b)\notag\\
    &+\tfrac{1}{2c^2\,\vert h\vert^2}{\cal L}_0(x,b)^2+\ldots\Big]\Bigg)\:.
\end{align}
Further expanding the exponential, and keeping only terms which produce a quadratic ation in the remaining fields, we find
\begin{align}
    \int_X e^\chi{\cal L}(x,b)&=i\,c\,\vert h\vert\vert\Omega\vert^2{\rm Vol}(X)+\int_X\left[{\cal L}_0(x,b)+\tfrac{i}{2c\,\vert h\vert}{\cal L}_0(x,b)^2\right]+\ldots\notag\\
    &=i\,c\,\vert h\vert\vert\vert\Omega\vert^2{\rm Vol}(X)-\tfrac12\int_X ( \bar\partial x +\partial b) \cdot(H)^{-1}\cdot (\bar\partial x +\partial b)\notag\\
    &+\tfrac{i}{2c\vert h\vert}\int_X{\cal L}_0(b)^2+\mathcal{O}(x^3,b^3)\:,
\end{align}
where again
\begin{equation}
    {\cal L}_0(b)=\tfrac{i}{\vert\Omega\vert^2}(\Omega\lrcorner\delb b)\:.
\end{equation}
We hence see that integrating out the dilaton from the full action or from the quadratic approximation results in the same action up to quadratic order. The difference enters in higher order terms. In particular, exactly integrating out the dilaton will produce higher order non-local interaction terms.

\subsection{Infinite Distance Limits}
 \label{sec:infdistance}
Having set up the effective actions, we can discuss infinite distance limits in the moduli space and check whether a tower of states becomes exponentially light, as we would expect from the swampland distance conjecture. To discuss such limits, recall that the K\"ahler potential for the complex structure moduli space is given in terms of the pre-potential $\mathcal{G}(\mu_i)$ via \cite{Candelas:1990rm}
\begin{align}
    \mathcal{K}=-\text{log}\left( 2(\bar{\mathcal{G}}-\mathcal{G})+ (\mu_i-\bar \mu_i)(\del_i \mathcal{G}+ \bar{\del}_i \bar{\mathcal{G}} )       \right)\, .
\end{align}
As an example, consider the mirror quintic threefold, which is a simple example with a single complex structure modulus. We take $H=u H_0$, where $H_0$ is a fixed complex structure and $u$ denotes the direction in complex structure moduli space for which we are taking the limit. 
At large complex structure, the pre-potential for the mirror quintic becomes
\begin{equation}
   \mathcal{G}(u)= \frac{5}{3!} u^3\:,
\end{equation}
such that the K\"ahler potential is
\begin{equation}
    \mathcal{K} = -  \log\left (\frac{5}{6}(u-\bar{u})^3\right)\, .
\end{equation} 
In radial coordinates $u=e^{i \phi} \lambda$ and neglecting the irrelevant $\phi$-dependent terms, we have
\begin{align}
   \mathcal{K} \propto  - \log (\lambda^3) \, ,
\end{align}
from which we can compute the K\"ahler metric
\begin{equation}
    K_{\lambda \lambda} =\frac{3}{\lambda^2}\:.
\end{equation} 
We then want to obtain flat coordinates $T$ for the kinetic terms, i.e.
\begin{align}
   K_{\lambda \lambda} \dd \lambda \dd \lambda = \dd T \dd T  \, ,
\end{align} 
which is provided by 
\begin{equation}
    \sqrt{ 3}   {\lambda}^{-1} \dd \lambda = \dd T\, , \quad   
\end{equation} 
from which we obtain the relation
\begin{equation}
    \lambda^{-1}=e^{-\frac{T}{\sqrt{3}}}\:.
\end{equation} 
Recall that the elliptic operator in the action (see for example \eqref{eq:SeffDilaton2}) behaves like $H^{-1} \propto \lambda^{-1}$, which we interpret as the mass term for the infinite tower of internal hermitian degrees of freedom associated to the elliptic operator. For $T \rightarrow \infty$ this tower becomes exponentially light as expected from the swampland distance conjecture. 

The discussion was simple enough for the one-parameter case and will also apply for generic directions in $\mathcal{O}(100)$ dimensional complex structure moduli spaces. However, there may be special directions in which the Yukawa couplings vanish. Note that we can infer the structure of the Yukawa couplings by the virtue of mirror symmetry \cite{Candelas:1990pi,
Candelas:1990rm}: 
At large complex structure, the Yukawa coupling in $X_3$ is identified with the triple intersection numbers $D_{ijk}$ in the 
mirror $X_3^*$, which can be evaluated as 
\begin{align}
    D_{ijk} = \int_{X_3^*} D_i \cdot D_j \cdot D_k \, ,
\end{align}
with $D_i$ being a basis of integral two-forms. Indeed, the infinite distance limits in complex structure moduli space of $X_3$ correspond to infinite distance limits in K\"ahler moduli spaces of the mirror $X_3^*$. 

Notably, for large numbers of K\"ahler moduli, it has been observed that the intersection matrix indeed becomes sparse. This structure is closely related to fibration structures of the threefold as discussed in the works by Koll\'ar, Oguiso, and Wilson ~\cite{Kollar:2012pv,Oguiso1993ONAF,Wilson1994}. Divisors $D$ with $D^3=0$ or $D^2 \cdot D_m=0$, for any effective divisor $D_m$ are associated with a $T^2$ or K3 fibration of the CY, respectively. Infinite distance limits that send the volume of $D$ to infinity can therefore be thought of as decompactification limits of the base of the fibrations~\cite{Grimm:2019bey}.

As an example, consider the bi-cubic CICY given by the configuration matrix
\begin{align}
 X^*_{3}:   \left[
    \begin{array}{c|c} \mathbbm{P}^2 & 3 \\ 
    \mathbbm{P}^2 & 3 
    \end{array}
    \right]
\end{align} 
with intersection numbers 
\begin{align}
    D_{111}= D_{222}=0\, ,\quad D_{112}= D_{122}=3\, .
\end{align} 
The two vanishing self-intersections show the presence of two torus fibrations in $X_3^*$ over a $\mathbbm{P}^2$ base. The pre-potential for the two complex structures in the mirror bi-cubic $X_3$ in the large complex structure limit is therefore given by 
\begin{align}
    \mathcal{G}= \frac12\left( \mu_1 \mu_2^2 + \mu_2 \mu_1^2 \right) \, .
\end{align}
The problematic limits with vanishing Yukawa couplings are those that send one of the $\mu_i \rightarrow \infty$ and correspond to decompactifications of either $\mathbbm{P}^2$ bases in the two $T^2$ fibrations in the dual $X_3^*$. A general linear combination $\lambda_1\mu_1 +\lambda_2 \mu_2$ has a non-vanishing Yukawa interaction. For example along the diagonal direction $\lambda_1=\lambda_2=\lambda$, the pre-potential is $\mathcal{G}\propto \lambda^3$, which behaves in the same way as in the mirror quintic example.

The connection between infinite distance limits and fibration structures in the K\"ahler moduli has been explored in ~\cite{Grimm:2019bey}. 
The three principal types of infinite distance limits are denoted by type II, III and IV \cite{Grimm:2018cpv} and precisely correspond to sending the base of a K3 (or $T^4$), $T^2$ and the full threefold volume to infinite. Using our effective action, we have therefore identified the respective tower of states for the mirror dual of a type IV degeneration in the heterotic string
complex structure moduli space. It would be interesting to extend the discussion of our effective action to include degeneration limits in directions that correspond to type II and III limits in the K\"ahler sector of the mirror dual geometry.

\section{Elliptic complexes and symplectic cohomology}
\label{sec:elliptic}
In the above quadratic actions, we have pointed out putative gauge symmetries. In order to make this more rigorous, we check in this section that these structures indeed form part of elliptic complexes, implying that the corresponding cohomologies become finite-dimensional on compact geometries. This is also important when quantizing the respective actions later, since the cohomologies count the ``on-shell'' degrees of freedom, which ought to be finite and can (modulo potential complications from anomalies) be decoupled from the quantization procedure. Note also that the more formal BV-quantization procedure requires that the complexes involved be elliptic, which then allows one to write the (one-loop) partition function in terms of determinants of elliptic operators. 

We will show that the proposed naive gauge transformations are not quite correct, and in understanding the correct gauge complex we define a new symplectic cohomology associated to the background complex structure deformation. We illustrate this for the quadratic effective action~\eqref{eq:effect4} that includes the dilaton. We refer to this as the ``dilaton effective action'', though the dilaton does not appear explicitly in this quadratic action.

\subsection{The first term of the dilaton effective action}
\label{sec:4Dilaton1}
Let us begin by considering the first term of the dilaton effective action \eqref{eq:effect4},
\begin{equation}
    S_\text{eff}=\int_X \del x \star_H \del x\:.
\label{eq:effect4a}
\end{equation}
Here the kinetic operator in question takes the form
\begin{equation}
    D=\del \star_H \del\:.
\end{equation}
This operator forms part of a natural differential complex
\begin{equation}
    \label{eq:NaiveComplex}    0\rightarrow\Omega^{(0,1)}\xrightarrow{\del}\Omega^{(1,1)}\xrightarrow{D}\Omega^{(2,2)}\xrightarrow{\del}\Omega^{(3,2)}\rightarrow0\:.
\end{equation}
To check that this complex is elliptic, we replace $\del$ with $\xi\in\Omega^{(1,0)}$, the holomorphic part of a real one-form,
\begin{equation}
    \label{eq:NaiveComplexXi}    0\rightarrow\Omega^{(0,1)}\xrightarrow{\xi}\Omega^{(1,1)}\xrightarrow{D_\xi}\Omega^{(2,2)}\xrightarrow{\xi}\Omega^{(3,2)}\rightarrow0\:.
\end{equation}
If the new complex is exact for all non-vanishing $\xi$, i.e., the kernel at one stage is isomorphic to the image at the previous stage, the complex is elliptic. As is, the complex~\eqref{eq:NaiveComplex} appears not to be elliptic, as we explain next.

It is easy to see that the complex is exact at the first stage and last stage, as these stages form part of elliptic Dolbeault complexes. Consider therefore the stage $\Omega^{(1,1)}$. Clearly ${\rm ker}(\xi)\subseteq{\rm Im}(D_\xi)$. To show the converse, assume that $\alpha\in\Omega^{(1,1)}$ sits in ${\rm ker}(D_\xi)$. That is
\begin{equation}
    D_\xi\alpha=\xi\star_H (\xi\wedge\alpha)=0\:.
\end{equation}
From ellipticity of the $\del$-complex, it follows that
\begin{equation}
\label{eq:NotElliptic}
    \star_H(\xi\wedge\alpha)=\xi\wedge\beta
\end{equation}
for some $\beta\in\Omega^{(0,2)}$. However, the space of two-forms that are not in the kernel of $\xi$ is (point-wise) six-dimensional, and the operator $\star_H\circ\xi$ is an isomorphism on this space. However, since $\beta$ is a (0,2)-form, it lives in a (point-wise) three-dimensional subspace. Hence, only those $\alpha$ that map into this three-dimensional subspace can satisfy equation \eqref{eq:NotElliptic}. As we shall see, this subspace is related to $\alpha$ being (non-)primitive in an appropriate sense. This however leaves a potential three-dimensional sub-space of $\alpha$'s which can satisfy \eqref{eq:NotElliptic}, i.e. not in $\ker(\star_H\circ\xi)=\ker(\xi)={\rm Im}(\xi)$, implying that the complex \eqref{eq:NaiveComplex} is not elliptic. 

To define the primitivity notion we are interested in, we first define a new double complex\footnote{A similar complex appeared recently in \cite{Bonezzi:2024dlv}, in the context of the double copy of Chern-Simons theory and Kodaira-Spencer theory.}
\begin{equation}
\hat\Omega^{(p,q)}:=\Omega^{(p,0)}\otimes\wedge^qT^{(0,1)}\:.
\end{equation}
A normal form in $\Omega^{(p,q)}$ can be mapped to a form in $\hat\Omega^{(p,3-q)}$ using $\Omega^{\bar a\bar b\bar c}$. In particular, for our field we have $x\in\hat\Omega^{(1,2)}$. Note also that $h^{-1}\in\hat\Omega^{(1,1)}$, and in this sense it takes the role of a (complex) K\"ahler form. We use this form to define a notion of primitivity. We can then decompose
\begin{equation}
    x=x_p+h^{-1}\wedge v\:,
\end{equation}
where $x_p\in\hat{\cal P}^{(1,2)}\subset \hat\Omega^{(1,2)}$ is primitive and $v\in\hat\Omega^{(0,1)}$, and the wedge product is on $\hat\Omega^{(p,q)}$. Similar decompositions hold for forms of other degrees, just like ordinary primitivity in symplectic geometry. 

The operators $\del$ and $\delb$ can also be interpreted as acting on $\hat{\Omega}^{(p,q)}$. Note that $\del$ has degree $(1,0)$ whereas $\delb$ has degree $(0,-1)$. To see how they act, let $\alpha \in \hat{\Omega}^{(p,q)}$, 
\begin{align}
\alpha= \tfrac{1}{p!}\tfrac{1}{q!} \alpha_{\mu_1...\mu_p}{\,^{\bar \nu_1 ,...,\bar \nu_q}}\: \dd z^{\mu_1...\mu_p}{\,}_{\bar \nu_1 ,...,\bar \nu_q}\:, 
\end{align}
where 
\begin{align}
\dd z^{\mu_1...\mu_p}{\,}_{\bar \nu_1 ,...,\bar \nu_q}=\dd z^{\mu_1}\wedge...\wedge\dd z^{\mu_p}\otimes \dd z_{{\bar \nu_1}}\wedge...\wedge\dd z_{{\bar \nu_q}}\:,
\end{align}
and we use the metric to lower the indices. Then
\begin{align}
    \del \alpha &= \tfrac{1}{p!}\tfrac{1}{q!} \del_{\mu}\alpha_{\mu_1...\mu_p}{\,^{\bar \nu_1 ,...,\bar \nu_q}} \dd z^{\mu \mu_1...\mu_p}{\,_{\bar \nu_1...\bar \nu_q}}\:,\\
    \delb \alpha &= \tfrac{1}{p!}\tfrac{1}{q!} (-1)^{p+q+1}\nabla_{\bar{\nu}_1}\alpha_{\mu_1...\mu_p}{\,^{\bar \nu_1 ,...,\bar \nu_q}} \dd z^{\mu_1...\mu_p}{\,_{\bar \nu_2...\bar \nu_q}}\:,
    \label{htdelb}
\end{align}
where $\nabla$ is the connection defined in Appendix \ref{app:Geometry}.

Inspired by previous works in symplectic cohomology \cite{Tseng:2009gr, Tseng:2010kt, Tsai:2014ela} and their applications in physics \cite{Tseng:2011gv, Lau:2014fia, Marchesano:2014iea, Gray:2018kss, bedulli2023syz, Kupka:2024rvl}, we can decompose $\del$ as
\begin{equation}
    \del=\del_++h^{-1}\del_-\,,
\end{equation}
where $\del_+:\hat{\cal P}^{(p,q)}\rightarrow\hat{\cal P}^{(p+1,q)}$ and $\del_-:\hat{\cal P}^{(p,q)}\rightarrow\hat{\cal P}^{(p,q-1)}$. Consider then
\begin{equation}
    0=\del^2=\del_+^2+h^{-1}(\del_-\del_++\del_+\del_-)+h^{-1}h^{-1}\del_-^2\:.
\end{equation}
From the primitive decomposition, we get that each term vanishes separately,
\begin{equation}
\label{eq:PrimNilpotent}
    \del_+^2=0\,,\qquad h^{-1}(\del_-\del_++\del_+\del_-)=0\,,\qquad \del_-^2=0\,,
\end{equation}
similar to ordinary symplectic geometry. Note that for forms in $\hat{\cal P}^{(p,q)}$ of degree $p+q=3$, the middle equality is trivial, while for forms where $p+q<3$ it implies that $\del_+$ and $\del_-$ anti-commute. There are no primitive forms of total degree larger than $3$.

In terms of the primitive decomposition, the action becomes
\begin{equation}
    S=\int_X(h^{-1}\del_-x_p+h^{-1}\del_+v+h^{-1}h^{-1}\del_-v)\star_H(h^{-1}\del_-x_p+h^{-1}\del_+v+h^{-1}h^{-1}\del_-v)\:.
\end{equation}
Note that the Hodge-dual like isomorphism $\star_H$ is constructed with $h$ and its inverse, which are singlets under the primitive decomposition. It follows that $\star_H$ can only change the form representation, but not the primitivity type of a form. Note also that forms of the same primitivity type can only pair to give a singlet in one way, which we will view as a top-form which can then be integrated in the normal way.\footnote{A top-form on $\hat\Omega^{(p,q)}$ is easily converted to a normal top-form using $\bar\Omega$. For example, for $\alpha\in\hat\Omega^{(3,3)}$, we can construct a top-form $\tfrac{1}{3!}\alpha^{\bar a\bar b\bar c}\bar\Omega_{\bar a\bar b\bar c}\wedge\bar\Omega\in\Omega^{(3,3)}\:.$}
The above action hence splits into two terms $S_1(x_p,v)$ and $S_2(v)$ where
\begin{align}
    S_1(x,v)&=\int_X(h^{-1}\del_-x_p+h^{-1}\del_+v)\star_H(h^{-1}\del_-x_p+h^{-1}\del_+v)\:.\notag\\
    &\propto
    \int_Xh^{-1}(\del_-x_p+\del_+v)(\del_-x_p+\del_+v)\:,
\end{align}
and
\begin{equation}
    S_2(v)=\int_X(h^{-1}h^{-1}\del_-v)\star_H(h^{-1}h^{-1}\del_-v)\propto\int_Xh^{-1}h^{-1}h^{-1}\del_-v\del_-v\:.
\end{equation}
Consider the cross-terms of $S_1(x,v)$ proportional to
\begin{equation}
    \int_Xh^{-1}\del_-x_p\del_+v=\int_Xh^{-1}\del_- x_p\del v=\int_X\del x_p\del v=0\:,
\end{equation}
where the first equality uses the primitive decomposition,  the second equality follows from $\del$-closure of $h^{-1}$ and primitivity of $x_p$, and the final expression vanishes since the integrand is a total derivative. 

The term in $S_1$ involving $v$ becomes
\begin{equation}
    \int_Xh^{-1}\del_+v\del_+v=\int_Xh^{-1}\del v\del_+v=\int_Xh^{-1}v\del\del_+v=-\int_Xh^{-1}h^{-1}v\del_+\del_-v\:.
\end{equation}
A similar computation shows that this is proportional to $S_2(v)$. Note that $\del_+={\cal O}(1)$ and $\del_-={\cal O}(H)$. If we then re-scale our fields $(x_p,v)$ so that they do not scale with $H$, and absorb the numerical prefactors, we can write the action in a simple form as
\begin{equation}
\label{eq:PrimAction1}
    S=\vert h\vert^{-1}\int_X x_p\del_+\del_-x_p+\vert h\vert^{-\tfrac13}\int_Xh^{-1}h^{-1}v\del_+\del_-v\:.
\end{equation}
Note that we do not expect the last term to vanish in general, since this term is derived from the original action \eqref{eq:effect4a} as the term involving the part of $\del x$ that is point-wise proportional to $\Omega(h)$.

Taking inspiration from \cite{Tseng:2010kt}, we expect the kinetic operator of the first term of \eqref{eq:PrimAction1} to form part of an elliptic complex
\begin{equation}
\label{eq:PrimComplex1}
    0\rightarrow\hat{\cal P}^{(0,2)}\xrightarrow{\del_+}\hat{\cal P}^{(1,2)}\xrightarrow{\del_+\del_-}\hat{\cal P}^{(2,1)}\xrightarrow{\del_-}\hat{\cal P}^{(2,0)}\rightarrow0\:,
\end{equation}
while the kinetic operator of the second term of \eqref{eq:PrimAction1} is part of an elliptic complex
\begin{equation}
    \label{eq:PrimComplex2}
    0\rightarrow\hat{\cal P}^{(0,3)}\xrightarrow{\del_-}\hat{\cal P}^{(0,2)}\xrightarrow{\del_-}\hat{\cal P}^{(0,1)}\xrightarrow{\del_+\del_-}\hat{\cal P}^{(1,0)}\xrightarrow{\del_+}\hat{\cal P}^{(2,0)}\xrightarrow{\del_+}\hat{\cal P}^{(3,0)}\rightarrow0\:.
\end{equation}
These are the complexes we need to use when we BV-quantize the action \eqref{eq:effect4a}. We prove that they are elliptic in Appendix \ref{app:elliptic}.

\subsection{The second term of the dilaton effective action}
\label{sec:4Dilaton2}
Regarding the second term of the dilaton effective action \eqref{eq:effect4},
\begin{equation}
    \label{eq:effect6}
    \int_X \delb x \star'_{H} \delb x\:,
\end{equation}
a similar analysis can be done for its kinetic operator $D'=\delb  \star'_{H} \delb$. This is the leading term of dilaton effective action as we approach the boundary of complex structure moduli space, when $H\rightarrow\infty$, and so should give the leading contribution to the partition function in this limit.  

The appropriate double complex is
\begin{equation}
\check\Omega^{(p,q)}:=\wedge^pT^{(1,0)}\otimes\Omega^{(0,q)}\:.
\end{equation}
The field $x \in \Omega^{(1,1)}$ is mapped to $\check\Omega^{(2,1)}$ by contracting with $\bar\Omega$, that is, $x_{a\bar b} \rightarrow {x}_{\bar b}^{bc}= x_{a\bar b} \bar{\Omega}^{abc}$. Clearly $h \in \check\Omega^{(1,1)}$, so we can use $h$ as a complexified K\"ahler form and define primitivity with respect to it. We have the following decompositions:
\begin{align}
    x=x_p + h \wedge v\:,\qquad \delb = \delb_+ + h \wedge \delb_-\:,
\end{align}
where $\delb_+:\check{\cal P}^{(p,q)}\rightarrow\check{\cal P}^{(p,q+1)}$ and $\delb_-:\check{\cal P}^{(p,q)}\rightarrow\check{\cal P}^{(p-1,q)}$.

The computation is now analogous to the previous discussion. We note that $\delb_+={\cal O}(1)$ and $\delb_-={\cal O}(H^{-1})$. If we again re-scale our fields $(x_p,v)$ so that they do not scale with $H$, and absorb numerical prefactors, the part of the action in~\eqref{eq:effect6} can then be written as
\begin{equation}
\label{eq:PrimAction1b}
   S= \vert h\vert^{-\tfrac13}\int_X x_p\delb_+\delb_-x_p+\vert h\vert^{-1}\int_Xh\, h\,  v\,\delb_+\delb_-v\:.
\end{equation}
The complexes governing the first and second terms are respectively
\begin{align}
    \label{eq:PrimComplex1b}
    0\rightarrow\check{\cal P}^{(2,0)}\xrightarrow{\delb_+}\check{\cal P}^{(2,1)}&\xrightarrow{\delb_+\delb_-}\check{\cal P}^{(1,2)}\xrightarrow{\delb_-}\check{\cal P}^{(0,2)}\rightarrow0\:,\\
    0\rightarrow\check{\cal P}^{(3,0)}\xrightarrow{\delb_-}\check{\cal P}^{(2,0)}\xrightarrow{\delb_-}\check{\cal P}^{(1,0)}&\xrightarrow{\delb_+\delb_-}\check{\cal P}^{(0,1)}\xrightarrow{\delb_+}\check{\cal P}^{(0,2)}\xrightarrow{\delb_+}\check{\cal P}^{(0,3)}\rightarrow0\:.
    \label{eq:PrimComplex2b}
\end{align}
These complexes are again elliptic, and the appropriate ones to use for quantizing the action.

In the above we have decomposed the field $x$ in two ways, either in terms of the $\hat\Omega^{(p,q)}$ complex for the action \eqref{eq:effect4a}, or in terms of the $\check\Omega^{(p,q)}$ complex for the above action \eqref{eq:effect6}. It is also useful to see how the action \eqref{eq:effect6} can be written in a form where we view the field $x$ as part of the $\hat\Omega^{(p,q)}$ complex instead. To do so, first note that the map from $\hat\Omega^{(p,q)}$ to $\check\Omega^{(p,q)}$ takes primitive forms to primitive forms. This has to be the case, since $h$, $h^{-1}$, $\Omega$ and $\bar\Omega$ are all singlets under the primitive decomposition. 

For example, primitivity for $x_p\in\hat\Omega^{(1,2)}$ implies
\begin{equation}
    h^a_{\bar b}{x_{p\,a}}^{\bar b\bar c}=0\:.
\end{equation}
Viewing $x_p$ as an ordinary $(1,1)$-form, this is equivalent to 
\begin{equation}
\label{eq:xPrim1}
    \Omega(h)\wedge x_p=0\:,
\end{equation}
where
\begin{equation}
    \Omega(h)=\tfrac12h^a\wedge\Omega_{abc}\,\dd z^{bc}
\end{equation}
is the $(2,1)$-form corresponding to $h$. Contracting \eqref{eq:xPrim1} with another factor of $h$ is an isomorphism from $(3,2)$-forms to $(2,3)$-forms, giving
\begin{equation}
\label{eq:xPrim2}
    \Omega(h,h)\wedge x_p=0\:,
\end{equation}
where
\begin{equation}
        \Omega(h,h)=\tfrac12h^ah^b\wedge\Omega_{abc}\,\dd z^{c}\:.
\end{equation}
This equation is equivalent to $x_p$ being primitive when viewed as an element of the $\check\Omega^{(p,q)}$ complex. 

We then define a new operator on the $\hat\Omega^{(p,q)}$ complex, denoted as
\begin{equation}
    \delta = h^{\bar a}\nabla_{\bar a}\:,
\end{equation}
where $\nabla_{\bar a}$ is the Chern-type connection defined in Appendix \ref{app:Geometry}. We use this connection, as it is ``metric'' with respect to $h$. Moreover, it is torsion-free,
so ordinary anti-commuting derivatives can be replaced with $\nabla$, just like in K\"ahler geometry. Note also that
\begin{equation}
\d\::\;\;\hat{\Omega}^{(p,q)}\rightarrow\hat{\Omega}^{(p+1,q)}\:,
\end{equation}
and
\begin{equation}
    \d^2=\dd z^{ab}h_a^{\bar a}h_{b}^{\bar b}\nabla_{\bar a}\nabla_{\bar b}=0\:.
\end{equation}
Just like for $\del$, we can define operators $\d_+$ and $\d_-$ via
\begin{equation}
    \d=\d_++h^{-1}\wedge\d_-\:.
\end{equation}
The operators $\d_\pm$ satisfy analogous identities to $\del_\pm$, and we can further decompose $\delb x$ and the action \eqref{eq:effect6} in terms of these operators. Noting that $\d_+={\cal O}(H^{-1})$, and $\d_-={\cal O}(1)$, and after again re-scaling the fields appropriately leads to the effective action
\begin{equation}
\label{eq:PrimAction1c}
   S= \int_X x_p\d_+\d_-x_p+\vert h\vert^{\tfrac23}\int_Xh^{-1}h^{-1} v\,\d_+\d_-v\:,
\end{equation}
where now $x_p$ and $v$ are fields viewed as taking values in $\hat\Omega^{(p,q)}$. This is the action we get from \eqref{eq:PrimAction1b} when we view $x$ as a field in the $\hat\Omega^{(p,q)}$ complex instead.

\subsection{The full dilaton effective action}
\label{sec:4Dilaton3}
Let us consider the dilaton effective action \eqref{eq:effect4} written out in the primitive decomposition of the $\hat\Omega^{(p,q)}$ complex. Combining \eqref{eq:PrimAction1} and \eqref{eq:PrimAction1c}, we get the action
\begin{equation}
\label{eq:PrimAction1d}
    S= {\vert h\vert}^{-1}\int_X x_p(\del_+\del_-+\vert h\vert\d_+\d_-)x_p+\vert h\vert^{-\tfrac13}\int_X h^{-1}h^{-1} v(\del_+\del_-+\vert h\vert\d_+\d_-)v\:.
\end{equation}
We are interested in (elliptic) complexes in which the kinetic operator $(\del_+\del_-+\vert h\vert\d_+\d_-)$ appears. We make an educated guesses for what these complexes look like, but proving that they are indeed elliptic is beyond the scope of the present paper.

Let us begin by considering the first part of the above action,
\begin{equation}
\label{eq:PrimAction1dx}
    S(x_p)= {\vert h\vert}^{-1}\int_X x_p(\del_+\del_-+\vert h\vert\d_+\d_-)x_p\:.
\end{equation}
It is clear that the gauge-transformation of $x_p$ will include deformations by terms that are both $\del_+$- and $\d_+$-exact, that is
\begin{equation}
    \label{eq:Gauge-x1}
    \delta x_p=\del_+\alpha=\d_+\beta\:,
\end{equation}
for $\alpha,\beta\in\hat{\cal P}^{(0,2)}$. If $\gamma$ is a primitive form of total degree $3$, then
\begin{equation}
    \d_+\del_-\gamma=\del_+\d_-\gamma\:,
\end{equation}
as we show below. A sub-set of the transformations \eqref{eq:Gauge-x1} are then of the form
\begin{equation}
    \label{eq:Gauge-x2}
    \delta x_p=\del_+\d_-\gamma\:,  
\end{equation}
for $\gamma\in\hat{\cal P}^{(0,3)}$. We therefore propose the following complex governing the differential operator of \eqref{eq:PrimAction1dx}:
\begin{equation}
    \label{eq:PrimCompd}
    0\rightarrow{\hat {\cal P}}^{(0,3)}\xrightarrow{\del_+\d_-}\hat{\cal P}^{(1,2)}\xrightarrow{\del_+\del_-+\vert h\vert\d_+\d_-}\hat{\cal P}^{(2,1)}\xrightarrow{\del_+\d_-}{\cal P}^{(3,0)}\rightarrow0\:.
\end{equation}
Checking that this complex is elliptic, and thus conforming that \eqref{eq:Gauge-x2} exhausts the gauge transformations of $x_p$, is quite cumbersome, and beyond the scope of this paper. However, we remark without proof that the corresponding symbol complex appears to be exact for generic $\xi_p\in T^*_p X\backslash\{0\}$, though perhaps not in general. 

Let us now show that $\del_+\d_-=\d_+\del_-$ on the primitive forms of interest. In particular, consider $x_p\in\hat{\cal P}^{(1,2)}$ of total degree $3$. Note first that
\begin{equation}
    \left(\d x_p\right)^{\bar b\bar c}=h^{\bar a}\nabla_{\bar a}x_p^{\bar b\bar c}=2\,h^{[\bar b}\nabla_{\bar a}x_p^{\vert \bar a\vert\bar c]}\:,
\end{equation}
where we have used primitivity of $x_p$, i.e. $h^{-1}\wedge x_p=0$. Using equation \eqref{htdelb}, contraction of $\nabla_{\bar a}$ with $x_p$ is just the action of $\delb$. It follows that 
\begin{equation}
    \delta x_p=-h^{-1}\wedge(\delb x_p)\:.
\end{equation}
Let $\pi$ denote the projection onto primitive forms, and consider
\begin{equation}
    \del_+\d_-x_p=-\pi(\del\delb x_p)=\pi(\delb\del x_p)=\pi\left(\delb(h^{-1}\del_- x_p)\right)\:.
\end{equation}
Next, note that 
\begin{equation}
    \delb(h^{-1}\del_- x_p)^{\bar b}=-2\,\nabla_{\bar a}\left(-h^{[\bar a}\del_- x_p^{\bar b]}\right)=\d(\del_- x_p)^{\bar b}-h^{\bar b}\nabla_{\bar a}(\del_- x_p)^{\bar a}\:.
\end{equation}
The last term drops out when we do the primitive projection. We therefore get
\begin{equation}
    \del_+\d_-x_p=\pi\left(\d\del_- x_p)\right)=\d_+\del_-x_p\:,
\end{equation}
as desired. A similar computation shows $\del_+\d_-=\d_+\del_-$ on the other primitive forms of interest. 

Next, let us consider the second term of \eqref{eq:PrimAction1d}. To study the corresponding complex, it is convenient to define a new set of operators
\begin{align}
    \hat\dd&=\del+i\sqrt{\vert h\vert}\delta\\
    \hat\dd^c&=\del-i\sqrt{\vert h\vert}\delta\:.
\end{align}
Note that these operators are also nilpotent. For example,
\begin{equation}
    \hat\dd^2=i\sqrt{\vert h\vert}\left(\del\d+\d\del\right)=i\sqrt{\vert h\vert}\dd z^{ab}h_a^{\bar a}R_{\bar a b}=0\:,
\end{equation}
where $R$ is the curvature of $\nabla$. This vanishes, as the curvature is primitive as shown in Appendix \ref{app:Geometry}. As above, we decompose $\hat\dd$ and $\hat\dd_c$ as
\begin{align}
    \hat\dd&=\hat\dd_++h^{-1}\hat\dd_-\\
    \hat\dd^c&=\hat\dd^c_++h^{-1}\hat\dd^c_-    
\end{align}
where again $\hat\dd_\pm$ and $\hat\dd^c_\pm$ satisfies similar identities to $\del_\pm$ and $\d_\pm$. Let us compute\footnote{The first term of \eqref{eq:PrimAction1d} can similarly be written as
\begin{equation}
    S(x_p)={\vert h\vert}^{-1}\int_X x_p\,\hat\dd_+\hat\dd^c_- x_p\:.
\end{equation}}
\begin{align}
    S(v)=\int_X h^{-1}h^{-1} v\,\hat\dd_+\hat\dd^c_-v&=\int_X h^{-1}h^{-1} v(\del_+\del_-+\vert h\vert\d_+\d_-)v\notag\\
    &+i\sqrt{\vert h\vert}\int_Xh^{-1}h^{-1}v\,(\delta_+\del_--\del_+\delta_-)v\:. 
\end{align}
The last term vanishes upon integrations by parts. We therefore propose that $v$ has the gauge transformation
\begin{equation}
\label{eq:Gauge-v}
    \delta v=\hat\dd^c_-\kappa\:,
\end{equation}
for $\kappa\in\check{\cal P}^{(2,0)}$, with an associated differential complex
\begin{equation}
    \label{eq:PrimCompe}
    0\rightarrow\check{\cal P}^{(3,0)}\xrightarrow{\hat\dd^c_-}\check{\cal P}^{(2,0)}\xrightarrow{\hat\dd^c_-}\check{\cal P}^{(1,0)}\xrightarrow{\hat\dd_+\hat\dd^c_-}\check{\cal P}^{(0,1)}\xrightarrow{\hat\dd_+}\check{\cal P}^{(0,2)}\xrightarrow{\hat\dd_+}\check{\cal P}^{(0,3)}\rightarrow0
\end{equation}
governing the action $S(v)$. Again, we leave showing that this complex is elliptic, and thus confirming that \eqref{eq:Gauge-v} indeed spans the full set of gauge transformations, for future work. To investigate this question, one can expand the fields in an order parameter related to $H^{-1}$ and study the resulting actions order by order.

\section{The SYZ fibration and a three-dimensional toy model}
\label{sec:ToyModel}
According to the SYZ conjecture \cite{Strominger:1996it}, a Calabi-Yau manifold has a $T^3$ fibration over a topological three-sphere, and mirror symmetry corresponds to T-duality along the torus fiber. As one approaches the large complex structure limit, the three-torus of the SYZ-fibration shrinks while the topological three-sphere base of the fibration grows. A light tower of states could then arise either from winding string modes of the SYZ fiber or from KK modes of the SYZ base.

Ignoring complications concerning details of the degeneration loci of the fibration, we expect a six-dimensional theory to undergo an effective dimensional reduction to the three-sphere. This local three-dimensional description has been utilized to gain valuable insights into the nature of both heterotic and M-theory compactifications, see e.g.~\cite{Acharya:2001gy, Pantev:2009de, Braun:2018vhk, Barbosa:2019bgh, Barbosa:2019hts, Cvetic:2020piw, Acharya:2020xgn, Hubner:2020yde, Acharya:2021rvh, Acharya:2023xlx, Acharya:2023syl}, and~\cite{Alvarez-Garcia:2021pxo} for a discussion in the context of the emergent string conjecture. In this section, we make an educated guess for a similar reduction of our theory~\eqref{eq:SupAction}.

\subsection{The model}
Before we present the model, we need to define some terms. Let $\hat{\Omega}^{(p,q)}:= \Omega^p( S^3, \Omega^q (S^3) )$ be the space of $p$-forms on $S^3$ with values on $\Omega^q (S^3)$, not to be confused with the hatted double complex on $X$ considered above. This models an effective Dolbeault complex in which the fields are valued. An element $\alpha \in \hat{\Omega}^{(p,q)}$ can be written in some coordinate basis as
\begin{equation}
    \label{eq:AlphaDefinition}
    \alpha = \tfrac{1}{p!}\tfrac{1}{q!}\alpha_{a_1...a_p,m_1...m_q} \dd x^{a_1} \wedge \cdots \wedge \dd x^{a_p} \otimes \dd x^{m_1} \wedge \cdots \wedge \dd x^{m_q}\:,
\end{equation}
To shorten the expressions, we define the shorthand $\dd x^{a_1 \cdots a_p}$ for $\dd x^{a_1} \wedge \cdots \wedge \dd x^{a_p}$. We also need differentials to play the role of Dolbeault operators. As the tangent bundle of $S^3$ is topologically trivial, we could just set the connections on the tangent bundle to zero, and simply use the ordinary derivative to raise the $p$ and $q$ index. This does, however, not play well with the reduced geometric structure, which we model by the round metric on $S^3$. Indeed, recall that $h^{ab}$ is a symmetric invertible matrix, whose reduction is most conveniently modelled by the round metric.\footnote{Note that in the SYZ case, the metric on $S^3$ would not be the round metric but the pullback of the Calabi-Yau metric to $S^3$.}

In Appendix \ref{app:S3connections} we define two nilpotent operators
\begin{align}
    \dd_+\:&:\;\; \hat{\Omega}^{(p,q)} \to \hat{\Omega}^{(p+1,q)}\\
    \dd_-\:&:\;\; \hat{\Omega}^{(p,q)} \to \hat{\Omega}^{(p,q+1)}\:,
\end{align}
which also anti-commute
\begin{equation}
    \{\dd_+,\dd_-\}=\dd_+\dd_-+\dd_-\dd_+=0\:,
\end{equation}
so we have a double complex. These operators have the advantage that they are metric in an appropriate sense. Indeed, they are defined using flat metric connections on the tangent bundle. 

Inspired by the action \eqref{eq:SupAction}, we then consider the action
\begin{equation}
    S= \int_{S^3} \left(x_a \dd_{-}\mu^a+(H_{ab}+\nabla_+ x_{ab})\mu^a\mu^b+\mu^a\nabla_{+a}b\right)\:.
    \label{eq:SupAction3}
\end{equation}
where the fields are $x \in \hat{\Omega}^{(1,1)}$, $\mu\in\hat\Omega^{(0,1)}(TS^3)$, $b \in \hat{\Omega}^{(0,2)}$, and $H \in \hat{\Omega}^{(2,1)}$. In this particular case, if $x=x_{am}\dd x^a\otimes\dd x^m$, then $\dd_-$ acts as 
\begin{equation}
 \dd_{-}\, x = -(\nabla_{n}^{LC} x_{am}+ \tensor{\e}{_n^l_a}x_{lm})\dd x^a\otimes\dd x^{nm}\:,  
\end{equation}
and for $b= \frac12 b_{mn}\dd x^{mn}\in\hat{\Omega}^{(0,2)}$, 
\begin{equation}
    \dd_+b= \frac12\left( \nabla_a^{LC} b_{mn}- \tensor{\e}{_a^l_m}b_{ln}-\tensor{\e}{_a^l_n}b_{ml} \right) \dd x^a \otimes\dd x^{mn}\:.
\end{equation}
where $\nabla^{LC}$ is just the Levi-Civita connection associated to the round metric on $S^3$. It is natural to take
\begin{equation}
    H_{abm}=h\,\e_{abm}\:,
\end{equation}
i.e., associated to the volume form of the growing $S^3$, where the complex structure $h$ defines the size. We then have
\begin{equation}
    H_{ab}\mu^a\mu^b=H_{abm}\mu^a_n\mu^b_s \dd x^{mns}=h\,\e_{abm}\e^{mns}\mu^a_n\mu^b_s \dd x^{123}\:.   
\end{equation}
With this, it is straight-forward to compute the inverse of the matrix
\begin{equation}
    {H_a}^{ n}{\,_b}^{s}:=h\,\e_{abm}\e^{mns}\:,
\end{equation}
which is given by
\begin{equation}
\label{eq:ExHInv}
   {(H^{-1})_n}^{ a}{\,_s}^{ b}=\tfrac1h (\tfrac12 \delta_n^a \delta_s^b - \delta_s^a \delta_n^b )\:. 
\end{equation}
Using this, we can consider toy models on $S^3$ of all the above effective actions. For example, integrating out $\mu$ in \eqref{eq:SupAction3} as in Section \ref{sec:EffTheory}, and considering the quadratic part of the action, one gets
\begin{equation}
\label{eq:S3effect3}
    S_\text{eff} = -\tfrac14\int_{S^3}\tr\left((H^{-1} \dd_+ x)^2\right)-\tfrac14\int_{S^3} ( \dd_- x +\dd_+ b)_a^{m} (H^{-1}){}_{m}{}^a{}_{n}{}^b (\dd_- x +\dd_+ b)_b^{n}\:,    
\end{equation}
where we integrate a function with respect to the normalized volume form of $S^3$,
\begin{equation}
\int_{S^3}=\int_{S^3}{\rm vol}(S^3)\:,\;\;\;{\rm vol}(S^3)=\tfrac{1}{3!}\epsilon_{abc}\dd x^{abc}\:.    
\end{equation}
Here
\begin{equation}
    \label{eq:Matrix-dxdb}
    (\dd_- x +\dd_+ b)^{m}_a= (\nabla_{+a}b_{ns}+\nabla_{-n}x_{as}) \e^{mns}\:,
\end{equation}
and $\dd_+x$ is the matrix
\begin{equation}
    {(\dd_+\,x)_a}^{ p}{\,_b}^{q}=(\nabla_{+[a}x_{b]s} \epsilon ^{spq})\:,
\end{equation}
and the trace is over the double indices. 

We can insert~\eqref{eq:ExHInv} into~\eqref{eq:S3effect3} to obtain
\begin{equation}
\label{eq:effect3da}
    S_\text{eff} = -\frac{9}{16h^2} \int_{S^3}\tr\left( (\dd_+ x )^2\right)-\frac{1}{4h}\int_{S^3} \left(\tfrac{1}{2} (\tr(\dd_-x+\dd_+b))^2 -  \tr((\dd_-x+\dd_+b)^2)\right) \:,   
\end{equation}
where the trace in the last two terms is over the matrix indices of $\dd_-x+\dd_+b$, viewed as a matrix in \eqref{eq:Matrix-dxdb}.

Similarly, the action analogous to (\ref{eq:effect4}), which was derived by also considering linear fluctuations in the dilaton, is given by
\begin{equation}
\label{eq:effect3db}
    S_\text{eff} =  \int_{S^3} \left(-\frac{9}{16 h^2} \tr\left( (\dd_+ x )^2\right)- \frac{1}{8h}\left( \tr(\dd_-x)\right)^2 + \frac{1}{4h} \tr\left((\dd_-x)^2\right)\right) \:.    
\end{equation}
It is also possible to define a primitive decomposition of the fields as in Section \ref{sec:elliptic}, and similarly decompose the above actions. In the given model, this decomposition can be straight-forwardly related to irreducible representations of $SO(3)$.

\subsection{An alternative model}
As mentioned above, given that the tangent bundles of $S^3$ are topologically trivial, we could also pick our connections $\nabla_\pm$ such that the differentials $\dd_\pm$ (in an appropriate gauge) simply act as ordinary derivatives, for example
\begin{equation}
 \dd_{-}\, x = -\partial_{n} x_{am}\dd x^a\otimes\dd x^{nm}\:,  
\end{equation}
where we do not rename the operators by slight abuse of notation. These connections are perhaps simpler to work with in explicit computations. However, the connections are no longer ``metric'', which can lead to complications in defining the Laplacians, which will be required to quantize these models.

\section{Including gauge fields}
\label{sec:gauge}
Let us discuss including the gauge sector in the action. In this case, the flux is corrected as\footnote{We have neglected the higher derivative correction from the gravitational Chern-Simons term.}
\begin{equation}
H=\dd B+\frac{\alpha'}{4}\omega_{CS}(A)\:,    
\end{equation}
where $\omega_{CS}(A)$ is the Chern-Simons three-form of $A$. With this, the superpotential also changes, see~\cite{Ashmore:2018ybe}. We first discuss the case where we neglect dilaton fluctuations, where the action~\eqref{eq:SupAction} receives a contribution
\begin{equation}
  \Delta W_{\alpha}= \frac{\alpha'}{4}\int _X \tr\left(\alpha \wedge \delb_A \alpha  + \tfrac23 \alpha \wedge \alpha \wedge \alpha - 2\,\mu^d \wedge F_d \wedge \alpha-\alpha \wedge \mu^d \wedge (\del_A \alpha)_d  \right) \wedge \Omega\:,   
\end{equation}
where $\alpha:= \delta A \in \Omega^{(0,1)}(\End(V))$ is a deformation of the gauge connection $A$.

Note that $\mu$ remains quadratic in the full action, so we can still integrate it out. The new effective theory takes the form
\begin{align}
    S_\text{eff} &=\tfrac12\int_X\tr\log\left(H+\partial x\right)\notag\\
    &-\tfrac12\int_X \left( \bar\partial x +\partial b-\tfrac{\alpha'}{4}\tr(2\,\alpha \wedge F-\alpha \del_A \alpha)\right)\cdot (H+\partial x)^{-1} \cdot\left(\bar\partial x +\partial b-\tfrac{\alpha'}{4}\tr(2\,\alpha \wedge F-\alpha \del_A \alpha)\right)\notag\\
    &+\frac{\alpha'}{4}\int _X \tr\left(\alpha \wedge \delb_A \alpha  + \tfrac23 \alpha \wedge \alpha \wedge \alpha \right) \wedge \Omega
    \:,
\end{align}
where $(H+\partial x)^{-1}$ is again the symmetric pairing \eqref{eq:Hinv} on $\Omega^{(1,0)}(T^{(0,1)})$, and
\begin{equation}
    \left(\bar\partial x +\partial b-\tfrac{\alpha'}{4}\tr(2\,\alpha \wedge F-\alpha \del_A \alpha)\right)_a^{\bar d}
    =\tfrac{i}{2\vert\Omega\vert^2}\left(\delb x + \del b-\tfrac{\alpha'}{4}\tr(2\,\alpha \wedge F-\alpha \del_A \alpha)\right)_{a\bar b\bar c}{\Omega}^{\bar b\bar c\bar d}\:,
\end{equation}
appropriately normalized so that it remains topological, i.e., metric independent. 

When the dilaton is included, the action (\ref{eq:SwithDilaton}) receives a contribution
\begin{align}
  \Delta W_{\alpha}= \frac{\alpha'}{4}\int _X e^{\chi} \tr\left(\alpha \wedge \delb_A \alpha  + \tfrac23 \alpha \wedge \alpha \wedge \alpha - 2\,\mu^d \wedge F_d \wedge \alpha-\alpha \wedge \mu^d \wedge (\del_A \alpha)_d  \right) \wedge \Omega   
\end{align}
After integrating out $\mu$, we get the effective action
\begin{align}
    S_\text{eff} &=\tfrac12\int_X\tr\log\left[e^{\chi}\left(H+\partial x\right)\right]\notag\\
    &-\tfrac12\int_Xe^{\chi}\left( \bar\partial x +\partial b+\tfrac{\alpha'}{4}\tr(-2\,\alpha \wedge F+\alpha \del_A \alpha)\right)\cdot (H+\partial x)^{-1} \cdot\left(\bar\partial x +\partial b+\tfrac{\alpha'}{4}\tr(-2\,\alpha \wedge F+\alpha \del_A \alpha)\right)\notag\\
    &+\frac{\alpha'}{4}\int _X e^{\chi}\tr\left(\alpha \wedge \delb_A \alpha  + \tfrac23 \alpha \wedge \alpha \wedge \alpha \right) \wedge \Omega+\int_Xe^\chi\bar{\partial}b \wedge \Omega
    \:.
\end{align}
As in Section \ref{sec:dilaton}, we can expand the action up to quadratic order in the fields. If again the axio-dilaton has a large mass scale, we can expand $\chi$ to linear order and integrate it out to obtain $\delb b=0$. This implies on a Calabi-Yau manifold that $b$ is $\delb-$exact and can hence again be absorbed in $x$ by an appropriate field redefinition. 

The new quadratic effective action then takes the form
\begin{align}
    S_\text{eff} &= -\tfrac14\int_X\tr\left((H^{-1}\partial x)^2\right)- \tfrac12\int_X\left ( \bar\partial x -\tfrac{\alpha'}{2}\tr(\alpha \wedge F)\right)\cdot H^{-1} \cdot\left(\bar\partial x -\tfrac{\alpha'}{2}\tr(\alpha \wedge F)\right)\notag\\
    &+\frac{\alpha'}{4}\int _X \tr\left(\alpha \wedge \delb_A \alpha   \right) \wedge \Omega\:.
\end{align}
We will not examine this action in much further detail here, except to note the induced couplings between the hermitian degrees of freedom $x$, and the gauge degrees of freedom $\alpha$. Note also that if we ignore the hermitian degrees of freedom, by setting $x$ to zero, we get an induced mass term for the internal gauge degrees of freedom. This mass term also comes down at a rate $\mathcal{O}(H^{-1})$ as we approach the boundary of complex structure moduli space.

\section{Conclusions and outlook}
\label{sec:conclusion}
In this work we have explored the recently proposed version of a heterotic Kodaira-Spencer like gravity \cite{Ashmore:2018ybe}, derived from fluctuations of the heterotic superpotential. Treating the heterotic superpotential as an effective quantum field theory, and including mathematically motivated (non-perturbative) corrections, leads to a way to study the heterotic moduli space directly. This is non-trivial since the heterotic Bianchi identity mixes K\"ahler, complex structure, and bundle moduli into a combined moduli space with coupled off-shell fluctuations in the effective action.

Using this framework, we explored infinite distance limits and compared them with expectations from the swampland distance conjecture \cite{Ooguri:2006in}: At large complex structure, we can integrate out off-shell fluctuations resulting in an effective theory for the residual degrees of freedom. We confirmed that at large distances in the complex structure moduli space, or at large background flux, we get the expected qualitative behavior of an effective tower of states becoming light at an exponential rate in geodesic distance.

The new effective actions are also interesting in their own right. In particular, the differentials governing the quadratic terms form a part of new elliptic symplectic type differential complexes, where the background complex structure Beltrami differential plays the role of the symplectic form. It would be very interesting to study the mathematical properties of these complexes further, and to investigate the quantum properties of these new effective actions, such as the one-loop partition function, anomalies, etc. In particular, the full one-loop function has been computed in \cite{Ashmore:2023vji}, where it was studied in great detail without background fluxes turned on. Turning on background fluxes and torsion complicates this analysis, and our method of partially integrating out fields in the presence of fluxes could be useful in this regard.

It should also be noted that in order to integrate out the complex structure moduli, we relied on the background complex structure having a non-vanishing Yukawa coupling. The Yukawa couplings are related to the triple intersection numbers in the mirror dual geometry, which becomes sparse in a basis of K\"ahler cone generators when the numbers of K\"ahler moduli becomes large. The vanishing entries in the intersection matrix of the mirror are related to certain fibration structures, which become more abundant with increasing number of K\"ahler moduli. In fact, there is a direct connection to infinite distance limits in the K\"ahler moduli space in which only sub-varieties of a fibration degenerate and the rate at which towers become light \cite{Grimm:2018cpv,Grimm:2019bey}. The specific directions in the complex structure moduli space
that we can probe in this work correspond to the most drastic decompactification limits in the mirror dual K\"ahler moduli spaces, denoted as type IV in the nomenclature of \cite{Grimm:2018cpv}. It would be interesting to extend our scenario to other limits and identify the respective towers that become light. 

Motivated by the SYZ conjecture, we also proposed toy-models on $S^3$ for the given effective actions. As the $T^3$ shrinks and the $S^3$ grows at large complex structure distance, one might expect an effective dimensional reduction of the theory. Exploiting this, we propose an ansatz for such a reduced theory. These theories have the advantage that the geometry is explicit, leading to the potential to perform explicit computations, for example of the one-loop partition functions. Such computations often lead to topological invariants that characterize the parent heterotic and Calabi-Yau geometries. Of course, the interesting topological properties of the SYZ fibration come from its degeneration loci, which can potentially be modeled as topological defects in the reduced theory, such as analogs of Wilson lines. Understanding how to model such effects and their potential computational power is an interesting future direction.   

Throughout most of the paper we have set $\alpha'=0$, effectively decoupling the gauge degrees of freedom. In Section \ref{sec:gauge}, we briefly considered turning on $\alpha'$ to include these modes. It would be interesting to study such couplings further. In doing so, one needs to take into account other non-perturbative effects, such as world-sheet instantons. These corrections introduce a complex structure dependence of the superpotential via Pfaffian prefactors of the instanton corrections.  These might play an important role in our theories, especially when the tree-level Yukawa coupling vanishes. However, whether these corrections occur or are absent due to Beasley-Witten cancellation~\cite{Beasley:2003fx} also depends on the gauge sector of the theory~\cite{Buchbinder:2019hyb,Buchbinder:2019eal} and the compactness of the instanton moduli space, which we mostly ignored in this paper. On the other hand, string theory lore suggests that couplings of such effects and the off-shell modes are suppressed, given the vanishing of world-sheet instantons at higher genus~\cite{Dine:1986zy,Dine:1987bq}. It would be interesting to extend our analysis to such cases in the future.

From a mathematical point of view, turning on the gauge degrees of freedom in the superpotential leads to a theory which effectively couples Donaldson-Thomas theory \cite{thomas1997gauge, donaldson1998gauge} to geometric degrees of freedom (the metric and complex structure). The procedure of integrating out geometric modes as outlined in this paper effectively couples Donaldson-Thomas theory to background moduli, and the remaining geometric modes. It is interesting to study how this affects the properties of Donaldson-Thomas invariants, wall-crossing phenomena, and others. On a more speculative level, the modes coming down at infinite distance in complex structure moduli space should also play a role in understanding the physics behind moduli space compactification. 

Finally, the procedure of integrating out modes at large distance as outlined here can also be employed in other theories, such as heterotic $G_2$ compactifications and M-Theory/F-theory models. In particular, the F-theory superpotential is very similar to the heterotic superpotential, and it would be very interesting to investigate similar avenues in these settings. 

\section*{Acknowledgments}
We thank Anthony Ashmore, Jim Halverson, Aleksi Kurkela, David McNutt, Jock McOrist, Ruben Minasian, Savdeep Sethi, Charles Stickland-Constable, and David Tennyson for many interesting discussions. We particularly thank David McNutt and David Tennyson for their involvement at an early stage of this project. ES would like to thank the mathematical research institute MATRIX in Australia, where part of this research was performed, for a lively and rewarding research environment. The work of FR is supported by the NSF grants PHY-2210333 and PHY-2019786 (The NSF AI Institute for Artificial Intelligence and Fundamental Interactions). The work of FR and PKO is also supported by startup funding from Northeastern University. PKO would like to thank the KITP and the program ”What is String Theory? Weaving Perspectives Together” during the completion of this work. This research was supported in part by grant NSF PHY-2309135 to the Kavli Institute for Theoretical Physics (KITP).

\appendix
\section{Contributions to the superpotential from Condensates}
\label{app:Condensates}
In this appendix we motivate how the superpotential term in \eqref{eq:ShiftDeltaW} could arise physically, from gaugino condensation \cite{Dine:1985rz, derendinger1985low}, or more general fermionic condensates. We repeat the term here for the readers convenience
\begin{equation}
\label{eq:Wc}
    \Delta W_{\rm condensate}=\int_X\frac{c}{3!} \mu^a\mu^b\mu^c\Omega_{abc}\wedge\Omega\:.
\end{equation}
A consistent inclusion of effects from the gauge sector, such as a gaugino condensation, requires a full treatment of the theory including $\alpha'$-corrections, which is beyond the scope of the paper.

The authors of \cite{LopesCardoso:2003sp} argue that a gaugino condensate effectively shifts the superpotential to 
\begin{equation}
    \Delta W\rightarrow \Delta  W+\int_X\Sigma\wedge\Omega\:,
    \label{eq:WCondensate}
\end{equation}
where $\Sigma_{mnp}=\langle{\rm Tr}\left(\bar\chi\Gamma_{mnp}\chi\right)\rangle$ is the gaugino condensate. On-shell, when the gaugino $\chi$ satisfies the equation of motion, $\Sigma$ takes the form 
\begin{equation}
    \Sigma=\Lambda^3\,\Omega+\bar\Lambda^3\,\bar\Omega\:.
    \label{eq:OnShellCondensate}
\end{equation}
Here $\Lambda^3$ denotes the expectation value of the gaugino condensate in the four-dimensional spacetime. Similar terms also appear from dilatino condensates \cite{Manousselis:2005xa, Chatzistavrakidis:2012qb}, and potentially also gravitino condensates. 

Inserting \eqref{eq:OnShellCondensate} into \eqref{eq:WCondensate} produces a term of the form
\begin{equation}
   \int_X \bar\Lambda^3\,\bar\Omega\wedge\Omega\:.
\end{equation}
As it stands, this term is not holomorphic, which a superpotential term needs to be. Nor is it clear what the condensate term will look like for off-shell deformations, not satisfying the equations of motion. To remedy this, and with the field content at hand, one might anticipate a dependence of the condensate term on the (not necessarily on-shell) parameters of the form
\begin{equation}
\bar\Lambda^3\,\bar\Omega=c\,\Omega(\mu,\mu,\mu)=\frac{c}{3!}\mu^a\mu^b\mu^c\Omega_{abc}\:,
\end{equation}
for some complex number $c$, potentially depending on the axio-dilaton \cite{LopesCardoso:2003dvb}. This is precisely the form of the shift in \eqref{eq:Wc}, and we afford ourselves this replacement. 

We stress that the condensate analysis presented here is speculative and not a rigorous derivation of a physical effect that leads to the deformation \eqref{eq:Wc}. Nevertheless, the term is a mathematically motivated deformation that could arise in a physical theory.

\section{Connections and Curvature}
\label{app:Geometry}
Using the complex structure deformation $h\in\Omega^{(0,1)}(T^{(1,0)}X)$, which is a non-degenerate matrix if it has a non-vanishing Yukawa coupling, we can define a Chern-type connection on holomorphic indices
\begin{equation}
    \nabla_a\alpha_b={h_b}^{\bar c}\partial_a\left({h_{\bar c}}^d\alpha_d\right)\:,
\end{equation}
with curvature
\begin{equation}
    R{}_{\bar b a}{}^{c}{}_{d\phantom{\bar  b}\!\!}=\partial_{\bar b}\left({h_d}^{\bar c}\partial_a{h_{\bar c}}^c\right)\:,
\end{equation}
in close analogy with the curvature of the ordinary Chern connection. A similar connection can be defined on anti-holomorphic indices. Clearly the connection is ``metric'' with respect to~$h$. Furthermore, the connection also preserves the holomorphic top-form, since
\begin{equation}
    \nabla_a(\Omega_{bcd})={h_{b}}^{\bar b}{h_{c}}^{\bar c}{h_{d}}^{\bar d}\partial_a\left({h_{\bar b}}^{b}{h_{\bar c}}^{c}{h_{\bar d}}^{d}\Omega_{bcd}\right)\propto\vert h\vert{h_{b}}^{\bar b}{h_{c}}^{\bar c}{h_{d}}^{\bar d}\partial_a(\bar\Omega_{\bar b \bar c \bar d})=0\:,
\end{equation}
where we have used that ${h_{\bar b}}^{b}{h_{\bar c}}^{c}{h_{\bar d}}^{d}\Omega_{bcd}\propto\vert h\vert\Omega_{\bar b \bar c \bar d}$, and $\vert h\vert$ is point-wise constant as noted in the main text. 

Let us discuss further properties of this connection and its curvature. We first note that the curvature is Ricci-flat,
\begin{equation}
    R{}_{\bar b a}{}^a{}_{d\phantom{\bar  b}\!\!}=\partial_{\bar b}\left({h_d}^{\bar c}\partial_a{h_{\bar c}}^a\right)=0\:,
\end{equation}
since $h$ is divergence-free $\partial_a{h_{\bar c}}^a=0$. Note also that the connection symbols 
\begin{equation}
    \Omega{}_a{}^c{}_d={h_d}^{\bar c}\partial_a{h_{\bar c}}^a
\end{equation}
are torsion-free,
\begin{equation}
    \Omega{}_{[a}{}^c{}_{d]}=0\:.
\end{equation}
Indeed, if ${h_{\bar c}}^a$ is harmonic, then so is its inverse. To see this, note that the inverse is proportional to
\begin{equation}
    {h_c}^{\bar c}\propto\Omega_{abc}{h_{\bar a}}^a{h_{\bar b}}^b\Omega^{\bar a\bar b\bar c}\:,
\end{equation}
which is harmonic \cite{Candelas:1990pi,Strominger:1990pd}. Indeed, note that
\begin{equation}
    {h_c}^{\bar c}{h_{\bar d}}^{c}\propto \Omega_{abc}{h_{\bar d}}^{c}{h_{\bar a}}^a{h_{\bar b}}^b\Omega^{\bar a\bar b\bar c}\propto \vert h\vert \bar\Omega_{\bar d\bar a\bar b}\Omega^{\bar a\bar b\bar c}\propto \delta_{\bar d}^{\bar c}\:,
\end{equation}
as $\vert h\vert$ is point-wise constant proportional to the Yukawa coupling of $h$. Here $\propto$ denotes equal up to irrelevant constant numerical factors. 

The curvature is also primitive in the sense of Section \ref{sec:elliptic}. We use $\Omega^{\bar a\bar b\bar c}$ to raise indices, and view
\begin{equation}
R\in\hat\Omega^{(1,2)}\left({\rm End}(T^{(1,0)})\right)\:.
\end{equation}
We then see that
\begin{equation}
    {h_{\bar a}}^{a}{R^{\bar a\bar b}}_{a}{}^{c}{}_{d\phantom{\bar  b}\!\!}=\Omega^{\bar a\bar b\bar c}\delb_{\bar c}\left({h_{\bar a}}^{a}{h_{a}}^{\bar d}\del_d{{h_{\bar d}}^c}\right)=\Omega^{\bar a\bar b\bar c}\delb_{\bar c}\left(\del_d{{h_{\bar a}}^c}\right)=0\:,
\end{equation}
where in the first equality we have used the symmetry of the connection symbol (torsion-free), and $\delb$-closure of $h$ in both equalities.

\section{Proving ellipticity of complexes}
\label{app:elliptic}
In this appendix we prove that \eqref{eq:PrimComplex1} and \eqref{eq:PrimComplex2} are elliptic. Let us repeat these complexes here for convenience
\begin{align}
    0\rightarrow\hat{\cal P}^{(0,2)}\xrightarrow{\del_+}\hat{\cal P}^{(1,2)}&\xrightarrow{\del_+\del_-}\hat{\cal P}^{(2,1)}\xrightarrow{\del_-}\hat{\cal P}^{(2,0)}\rightarrow0\:,
    \label{eq:PrimComplex1br}
    \\
    0\rightarrow\hat{\cal P}^{(0,3)}\xrightarrow{\del_-}\hat{\cal P}^{(0,2)}\xrightarrow{\del_-}\hat{\cal P}^{(0,1)}&\xrightarrow{\del_+\del_-}\hat{\cal P}^{(1,0)}\xrightarrow{\del_+}\hat{\cal P}^{(2,0)}\xrightarrow{\del_+}\hat{\cal P}^{(3,0)}\rightarrow0\:.
    \label{eq:PrimComplex2br}
\end{align}
Let us begin with the complex \eqref{eq:PrimComplex1br}. Using the middle relation of \eqref{eq:PrimNilpotent} and the comment below the equation,  it is easy to check that this is indeed a differential complex. To prove ellipticity, we again replace $\del$ with $\xi$ to get the complex
\begin{equation}
\label{eq:ExactPrimComplex1}
    0\rightarrow\hat{\cal P}^{(0,2)}\xrightarrow{\xi_+}\hat{\cal P}^{(1,2)}\xrightarrow{\xi_+\xi_-}\hat{\cal P}^{(2,1)}\xrightarrow{\xi_-}\hat{\cal P}^{(2,0)}\rightarrow0\:.
\end{equation}
Again, it is easy to check that this is a complex. 

To show exactness at the first level, we pick $\alpha\in\hat{\cal P}^{(0,2)}$ and insist that the primitive part of $\xi\wedge\alpha$ vanishes. Defining $\xi_{\bar a}=h_{\bar a}^a\xi_a$, this becomes equivalent to
\begin{equation}
    \xi_{(\bar a}\alpha_{\bar b)}=0\:,
\end{equation}
where we view $\alpha$ as an ordinary $(0,1)$-form. To see that this implies $\alpha=0$, we work point-wise and pick a frame where $(\xi_{\bar 1},\xi_{\bar 2},\xi_{\bar 3})=(1,0,0)$ (which is possible wlog since $\xi\neq0$).

Showing exactness at the second level is a bit more cumbersome. We pick $\beta\in\hat{\cal P}^{(1,2)}$. Being in the kernel of $\xi_+\xi_-$ means that $\xi\wedge\xi_-\beta$ is non-primitive, which is equivalent to
\begin{equation}
    \xi_{c}h_{\bar b}^{a}\xi_{[a}\beta_{b]}^{\bar b\bar c}\dd z^{cb}=\frac12h_{\bar b}^{a}\xi_{a}\xi_{[c}\beta_{b]}^{\bar b\bar c}\dd z^{cb}=h_{[c}^{\bar c}v_{b]}\dd z^{cb}
\end{equation}
for some $v_b$. This implies that
\begin{equation}
    \frac12h_{\bar c}^ch_{\bar b}^{a}\xi_{a}\xi_{[c}\beta_{b]}^{\bar b\bar c}=\frac14h_{\bar c}^ch_{\bar b}^{a}\xi_{a}\xi_{c}\beta_{b}^{\bar b\bar c}=h_{\bar c}^ch_{[c}^{\bar c}v_{b]}=\delta^c_{[c}v_{b]}=v_b\:.
\end{equation}
But the left-hand side vanishes as $\beta_{b}^{\bar b\bar c}$ is anti-symmetric in $\bar b\bar c$, which induces $\xi_{[a}\xi_{c]}=0$. It follows that
\begin{equation}
    \xi\wedge h_{\bar b}^{a}\xi_{[a}\beta_{b]}^{\bar b\bar c}\dd z^{b}=0\:,
\end{equation}
which implies
\begin{equation}
    h_{\bar b}^{a}\xi_{[a}\beta_{b]}^{\bar b\bar c}=\tfrac12h_{\bar b}^{a}\xi_{a}\beta_{b}^{\bar b\bar c}=\xi_b w^{\bar c}
\end{equation}
for some $w^{\bar c}$. Primitivity of $\beta$ implies that $h_{\bar c}^b\xi_bw^{\bar c}=0$. The above equation is then (modulo potentially re-scaling $w$) equivalent to
\begin{equation}
    \xi\wedge\beta=h^{-1}\wedge\xi\wedge w\:.
\end{equation}
That is
\begin{equation}
    \beta=h^{-1}\wedge w+\tilde\beta\:,
\end{equation}
where $\tilde\beta\in{\rm ker}(\xi)={\rm Im}(\xi)$. As $\beta$ is primitive, it follows that $\beta\in{\rm Im}(\xi_+)$, showing exactness at the second level. 

Next note the point-wise dimensions $\vert\hat{\cal P}^{(1,2)}\vert=\vert\hat{\cal P}^{(2,1)}\vert=6$. Injectivity of the first map, together with exactness at the second level implies $\vert {\rm Im}(\xi_+\xi_-)\vert =3$. Recall also that ${\rm Im}(\xi_+\xi_-)$ is a sub-space of ${\rm ker}(\xi_-)$. If we can show that the last map is onto (exactness at the last level), it follows that $\vert{\rm ker}(\xi_-)\vert=3$ and hence $\vert{\rm Im}(\xi_+\xi_-)\vert=\vert{\rm ker}(\xi_-)\vert$, so they are the same space, i.e., we have exactness at the third level as well. 

To show exactness at the last level, note that for $\gamma\in\hat{\cal P}^{(2,1)}$ we have
\begin{equation}
    (\xi_-\gamma)_{ab}=\xi_{\bar a}\gamma_{ab}^{\bar a}\:.
\end{equation}
Picking $\gamma$ of the form $\gamma_{ab}^{\bar a}=\kappa_{ab}u^{\bar a}$, it is clear that we can generate any form in $\hat{\cal P}^{(2,0)}$ as long as we pick $u$ so that $\xi_{\bar a}u^{\bar a}\neq 0$. This shows exactness at the last level, and hence the sequence \eqref{eq:ExactPrimComplex1} is exact. The complex \eqref{eq:PrimComplex1b} is therefore elliptic. 

Showing that the complex \eqref{eq:PrimComplex2b} is elliptic is more straightforward. The complex with $\del$ replaced by $\xi$ is
\begin{equation}
    \label{eq:Ell PrimComplex2}
    0\rightarrow\hat{\cal P}^{(0,3)}\xrightarrow{\xi_-}\hat{\cal P}^{(0,2)}\xrightarrow{\xi_-}\hat{\cal P}^{(0,1)}\xrightarrow{\xi_+\xi_-}\hat{\cal P}^{(1,0)}\xrightarrow{\xi_+}\hat{\cal P}^{(2,0)}\xrightarrow{\xi_+}\hat{\cal P}^{(3,0)}\rightarrow0\:.
\end{equation}
As $\xi_+=\xi$ on $\hat{\cal P}^{(p,0)}$, the complex is exact at the last two levels. A similar observation implies the complex is exact at the first two levels. At the third level, if $\alpha\in\hat{\cal P}^{(0,1)}$ and 
\begin{equation}
    \xi_+\xi_-\alpha=0\:,
\end{equation}
this can only happen if $\xi_-\alpha=0$, which implies $\alpha\in{\rm Im}(\xi_-)$ by exactness of the $\xi_-$-complex on $\hat{\cal P}^{(0,q)}$ forms. This is similar to how one argues ellipticity of the ordinary $\dd^\dagger$-complex and shows that the complex is exact at the third level. Finally, at the fourth level,  if $\beta\in\hat{\cal P}^{(1,0)}$ is in the kernel of $\xi_+$, then clearly it can be written as
\begin{equation}
    \beta=\xi_+(\xi_-v)
\end{equation}
for some appropriately chosen $v$ not in the kernel of $\xi_-$. This shows exactness at level four, and the complex \eqref{eq:PrimComplex2b} is elliptic.

\section{Connections on the three-sphere}
\label{app:S3connections}
In this appendix we define and analyze properties of the connections $\dd_\pm$ that we use on the double complex $\hat\Omega^{(p,q)}$ on $S^3$ in Section \ref{sec:ToyModel}.

\subsection*{Flat connections on the three-sphere}
Let $X= S^3$, and consider the following two connections on $X$:
\begin{align*}
    \dd_{\pm} = \dd + \Gamma^{LC} \pm \epsilon_{\pm}\:,
\end{align*}
where the action of $\epsilon_{\pm}$ is defined below in terms of the three-dimensional volume form. These differentials act on the double complex $\hat\Omega^{(p,q)}$ on $S^3$ as
\begin{align*}
    \dd_+\:&:\;\;\hat\Omega^{(p,q)}\rightarrow\hat\Omega^{(p+1,q)}\\
    \dd_-\:&:\;\;\hat\Omega^{(p,q)}\rightarrow\hat\Omega^{(p,q+1)}\:.
\end{align*}
Letting $\alpha\in\hat\Omega^{(p,q)}$ as in~\eqref{eq:AlphaDefinition}, these differentials act as
\begin{align*}
     \dd_+ \alpha=dx^a\nabla_a \alpha + \e_+\cdot \alpha\:,\qquad
      \dd_- \alpha=(-1)^{|\alpha|}\left(\nabla_m \alpha dx^m - \e_-\cdot \alpha\right)\:.
\end{align*}
where $|\alpha|:= p+q$ and $\nabla$ is the Levi-Civita connection for the round metric on $S^3$. Let us further define $\e_+ \cdot \alpha = \dd x^b (\e _b \cdot \alpha)$, where
\begin{equation}
    \epsilon_b \cdot \alpha = - \frac{1}{p!q!}\sum_{i=1}^q \tensor{\e}{_b^l_{m_i}} \alpha_{a_1\ldots a_p,m_1\ldots \widehat{m}_il\ldots m_q} \dd x^{a_1 \ldots a_p} \otimes \dd x^{m_1 \ldots m_q}\:.
\end{equation}
The hat on an index $\widehat{m}_i$ denotes omission of this index. Similarly, we define $\e_-\cdot \alpha =  (\e_n\cdot \alpha) \dd x^n$, where
\begin{equation}
\epsilon_n \cdot \alpha = - \frac{1}{p!q!}\sum_{i=1}^p \tensor{\e}{_n^l_{a_i}} \alpha_{a_1\ldots\widehat{a}_il \ldots a_p,m_1\ldots m_q} \dd x^{a_1 \ldots a_p} \otimes \dd x^{m_1 \ldots m_q}. 
\end{equation}

\bigskip

\textbf{Lemma:} These connections are flat on $S^3$. That is $\dd_+^2=\dd_-^2=0$.

\begin{proof}
We have that
\begin{equation}
    \dd_+^2 \alpha =  \dd x^{ba} \left( \nabla_b+\e_b )(\nabla_a  + \e_a  \right) \alpha  = \dd x^{ba}\left( \nabla_b \nabla_a+ \nabla_b \e_a  + \e_b \nabla_a +  \e_b \e_a \right) \alpha  
\end{equation}
It is easy to see that $\nabla_b $ and $\e_a$ commute. Moreover, from swapping $a$ and $b$ it is clear that $\dd x^{ba}(\nabla_b \e_a  +  \nabla_a\e_b)\alpha =0.$ Therefore
\begin{equation}
   \dd_+^2 \alpha = \dd x^{ba} \left( \frac12 [\nabla_b, \nabla_a] +  \e_b \e_a \right) \alpha \:.
\end{equation}
We have also that
\begin{align*}
  \dd x^{ba} \e_b \e_a  \alpha  &=-\e_b \cdot \left(\frac{1}{p!q!}\sum_{i=1}^q \tensor{\e}{_a^k_{m_i}} \alpha_{a_1\ldots a_p,m_1\ldots \widehat{m}_ik\ldots  m_q} \dd x^{baa_1 \ldots a_p} \otimes \dd x^{m_1 \ldots m_q}\right)   \\
   &=  \frac{1}{p!q!}\sum_{i=1}^q \tensor{\e}{_b^l_{m_i}} \tensor{\e}{_a^k_l}  \alpha_{a_1\ldots a_p,m_1\ldots  \widehat{m}_ik\ldots m_q} \dd x^{baa_1 \ldots a_p} \otimes \dd x^{m_1 \ldots m_q}+ \\
   &\phantom{=~} \frac{1}{p!q!}\sum_{i\neq j}^q \tensor{\e}{_b^l_{m_j}} \tensor{\e}{_a^k_{m_i}}  \alpha_{a_1\ldots a_p,m_1\ldots \widehat{m}_ik\ldots \widehat{m}_jl\ldots m_q} \dd x^{baa_1 \ldots a_p} \otimes \dd x^{m_1 \ldots m_q}\:.
\end{align*}
After swapping $a$ and $b$ and interchanging the roles of $i$ and $j$, it can be seen that the second term in the above expression vanishes. Therefore, 
\begin{align*}
\dd x^{ba}\e_b \e_a  \alpha  = \frac{1}{p!q!}\sum_{i=1}^q \delta_b^k g_{m_i a} \alpha_{a_1\ldots a_p,m_1\ldots \widehat{m}_ik\ldots m_q} \dd x^{ba a_1 \ldots a_p} \otimes \dd x^{m_1 \ldots m_q}\:.
\end{align*}
Recall that the Riemann tensor of the round metric for a sphere of the appropriate radius is given by
\begin{equation}\label{risp}
    R_{ijkl}= g_{ik}g_{jl}-g_{jk}g_{il}\:.
\end{equation}
So,
\begin{multline*}
\dd x^{ba}[\nabla_{b}, \nabla_{a}] \alpha  =- \bigg(\frac{1}{p!q!}\sum_{i=1}^{p} \tensor{R}{_b_a^l_{a_i} } \alpha_{a_1\ldots \widehat{a}_il\ldots a_p,m_1\ldots m_q}   \\+ \frac{1}{p!q!}\sum_{i=1}^{q} \tensor{R}{_b_a^l_{m_i} } \alpha_{a_1\ldots a_p,m_1\ldots \widehat{m}_il \ldots m_q}  \bigg)  \dd x^{baa_1 \cdots a_p} \otimes \dd x^{m_1 \cdots m_q}  \:.    
\end{multline*}
From \eqref{risp}, we know that $\tensor{R}{_b_a^l_{a_i}}=\delta_b^l g_{aa_i}-\delta_a^l g_{ba_i}$. Hence, after contracting with $\dd x^{baa_i}$ the first term in the equation above vanishes. Similarly, we get that $\tensor{R}{_b_a^l_{m_i}}=\delta_b^l g_{am_i}-\delta_a^l g_{bm_i}$, and after contracting with $\dd x^{ba}$ this gives rise to $2\delta_b^l g_{am_i} \dd x^{ba}$. We therefore have:
\begin{align*}
   \dd x^{ba} [\nabla_{b}, \nabla_{a}] \alpha  &=- 2 \sum_{i=1}^{q} \delta_b^l g_{am_i} \alpha_{a_1\ldots a_p,m_1\ldots  \widehat{m}_il\ldots m_q}  \dd x^{baa_1 \cdots a_p} \otimes \dd x^{m_1 \cdots m_q}  \\
    &=-2\dd x^{ba}\e_b \e_a  \alpha \:.
\end{align*}
This shows that $\dd_+^2=0$. An analogous computation shows that $\dd_-^2=0$.
\end{proof}

We can hence associate two nilpotent operators to the complex $\hat\Omega^{(p,q)}$. In order for it to be a double complex we must also have that $\dd_+$ and $\dd_-$ anti-commute.

\bigskip

\textbf{Lemma:}
\begin{align*}
    \{ \dd_+,\dd_- \}=0 
\end{align*}
\begin{proof}
We have
\begin{align}
\begin{split}
   \dd_+\dd_- \alpha &=  (-1)^{|\alpha|} \dd x^b( \nabla_b+\e_b )(\nabla_n  - \e_n  ) \alpha  \dd x^n \\
   &= (-1)^{|\alpha|}\dd x^b ( \nabla_b \nabla_n- \nabla_b \e_n  + \e_b \nabla_n -  \e_b \e_n) \alpha  \dd x^n \:.
\end{split}
\end{align}
It is easy to see that $\dd x^b\e_b \nabla_n \alpha  \dd x^n =\dd x^b( -\tensor{\e}{_b^k_n}\nabla_k  + \nabla_n \e_b) \alpha  \dd x^n$. So,
\begin{equation}
\dd_+\dd_- \alpha = (-1)^{|\alpha|}\dd x^b( \nabla_b \nabla_n- \nabla_b \e_n  -\tensor{\e}{_b^k_n}\nabla_k  + \nabla_n \e_b - \e_n \e_b) \alpha  \dd x^n.
\end{equation}
Similarly, we can see that
\begin{equation}
    \dd_-\dd_+ \alpha = (-1)^{|\alpha|+1}\dd x^b( \nabla_n \nabla_b+ \nabla_n \e_b  +\tensor{\e}{_n^k_b}\nabla_k  - \nabla_b \e_n -  \e_b \e_n) \alpha  \dd x^n \:.
\end{equation}
A straight-forward calculation shows that $\e_n \cdot \e_b - \e_b \cdot \e_n = \tensor{\e}{_n^c_b} \e _c -\tensor{\e}{_b^m_n} \e_m$, and hence
\begin{equation}\label{anti}
   \{ \dd_+,\dd_- \}= (-1)^{|\alpha|}\dd x^b(2 \nabla_{[b} \nabla_{n]}+\tensor{\e}{_n^c_b} \e _c -\tensor{\e}{_b^m_n} \e_m )\alpha \dd x^n. 
\end{equation}
Now, the components of $ \tensor{\e}{_n^c_b} \e _c \cdot \alpha$ are
\begin{equation}
    - \sum_{i=1}^q \tensor{\e}{_n^c_b}\tensor{\e}{_c^l_{m_i}} \alpha_{a_1\ldots a_p,m_1\ldots \widehat{m}_il \ldots m_q} =- \sum_{i=1}^q  (\delta^l_bg_{nm_i}-g_{bm_i}\delta^l_n) \alpha_{a_1\ldots a_p,m_1\ldots  \widehat{m}_il \ldots m_q}\:,
\end{equation}
and hence
\begin{equation}\label{ant1}
   \dd x^b \tensor{\e}{_n^c_b} \e _c \alpha  \dd x^n=  \sum_{i=1}^q  g_{bm_i}\delta^l_n \alpha_{a_1\ldots a_p,m_1\ldots  \widehat{m}_il \ldots m_q} \dd x^{ba_1 \cdots a_p} \otimes \dd x^{m_1 \cdots m_qn}\:,
\end{equation}
since $g_{nm_i}\dd x^{m_in}=0$. Similarly
\begin{equation}\label{ant2}
  \dd x^b\tensor{\e}{_b^m_n} \e _m \alpha  \dd x^n=  \sum_{i=1}^p  g_{na_i}\delta^l_b \alpha_{a_1\ldots \widehat{a}_il \ldots a_p,m_1\ldots  m_q} \dd x^{ba_1 \cdots a_p} \otimes \dd x^{m_1 \cdots m_qn}\:.  
\end{equation}
On the other hand
\begin{multline*}
 2 \dd x^b\nabla_{[b} \nabla_{n]} \alpha \dd x^n=- \bigg(\sum_{i=1}^{p} \tensor{R}{_n_b^l_{a_i} } \alpha_{a_1\ldots \widehat{a}_il \ldots a_p,m_1\ldots  m_q}  \\ + \sum_{i=1}^{q} \tensor{R}{_n_b^l_{m_i} } \alpha_{a_1\ldots a_p,m_1\ldots  \widehat{m}_il \ldots m_q}  \bigg) \dd x^{ba_1 \cdots a_p} \otimes \dd x^{m_1 \cdots m_qn}.    
\end{multline*}
It is easy to check that this term precisely cancels the terms \eqref{ant1} and \eqref{ant2}, so \eqref{anti} vanishes, obtaining the desired result.
\end{proof}

\bibliographystyle{JHEP}
\bibliography{refs}

\providecommand{\href}[2]{#2}\begingroup\raggedright\begin{thebibliography}{10}

\bibitem{Vafa:2005ui}
C.~Vafa, \emph{{The String landscape and the swampland}},  \href{http://arxiv.org/abs/hep-th/0509212}{{\tt hep-th/0509212}}.

\bibitem{vanBeest:2021lhn}
M.~van Beest, J.~Calder\'on-Infante, D.~Mirfendereski and I.~Valenzuela, \emph{{Lectures on the Swampland Program in String Compactifications}}, \href{http://dx.doi.org/10.1016/j.physrep.2022.09.002}{\emph{Phys. Rept.} {\bf 989} (2022) 1--50}, [\href{http://arxiv.org/abs/2102.01111}{{\tt 2102.01111}}].

\bibitem{Grana:2021zvf}
M.~Gra\~na and A.~Herr\'aez, \emph{{The Swampland Conjectures: A Bridge from Quantum Gravity to Particle Physics}}, \href{http://dx.doi.org/10.3390/universe7080273}{\emph{Universe} {\bf 7} (2021) 273}, [\href{http://arxiv.org/abs/2107.00087}{{\tt 2107.00087}}].

\bibitem{Agmon:2022thq}
N.~B. Agmon, A.~Bedroya, M.~J. Kang and C.~Vafa, \emph{{Lectures on the string landscape and the Swampland}},  \href{http://arxiv.org/abs/2212.06187}{{\tt 2212.06187}}.

\bibitem{Ooguri:2006in}
H.~Ooguri and C.~Vafa, \emph{{On the Geometry of the String Landscape and the Swampland}}, \href{http://dx.doi.org/10.1016/j.nuclphysb.2006.10.033}{\emph{Nucl. Phys. B} {\bf 766} (2007) 21--33}, [\href{http://arxiv.org/abs/hep-th/0605264}{{\tt hep-th/0605264}}].

\bibitem{Etheredge:2022opl}
M.~Etheredge, B.~Heidenreich, S.~Kaya, Y.~Qiu and T.~Rudelius, \emph{{Sharpening the Distance Conjecture in diverse dimensions}}, \href{http://dx.doi.org/10.1007/JHEP12(2022)114}{\emph{JHEP} {\bf 12} (2022) 114}, [\href{http://arxiv.org/abs/2206.04063}{{\tt 2206.04063}}].

\bibitem{Lee:2019wij}
S.-J. Lee, W.~Lerche and T.~Weigand, \emph{{Emergent strings from infinite distance limits}}, \href{http://dx.doi.org/10.1007/JHEP02(2022)190}{\emph{JHEP} {\bf 02} (2022) 190}, [\href{http://arxiv.org/abs/1910.01135}{{\tt 1910.01135}}].

\bibitem{Brodie:2021ain}
C.~R. Brodie, A.~Constantin, A.~Lukas and F.~Ruehle, \emph{{Swampland conjectures and infinite flop chains}}, \href{http://dx.doi.org/10.1103/PhysRevD.104.046008}{\emph{Phys. Rev. D} {\bf 104} (2021) 046008}, [\href{http://arxiv.org/abs/2104.03325}{{\tt 2104.03325}}].

\bibitem{Brodie:2021nit}
C.~R. Brodie, A.~Constantin, A.~Lukas and F.~Ruehle, \emph{{Geodesics in the extended K\"ahler cone of Calabi-Yau threefolds}}, \href{http://dx.doi.org/10.1007/JHEP03(2022)024}{\emph{JHEP} {\bf 03} (2022) 024}, [\href{http://arxiv.org/abs/2108.10323}{{\tt 2108.10323}}].

\bibitem{Lanza:2021udy}
S.~Lanza, F.~Marchesano, L.~Martucci and I.~Valenzuela, \emph{{The EFT stringy viewpoint on large distances}}, \href{http://dx.doi.org/10.1007/JHEP09(2021)197}{\emph{JHEP} {\bf 09} (2021) 197}, [\href{http://arxiv.org/abs/2104.05726}{{\tt 2104.05726}}].

\bibitem{Castellano:2023jjt}
A.~Castellano, I.~Ruiz and I.~Valenzuela, \emph{{Stringy Evidence for a Universal Pattern at Infinite Distance}},  \href{http://arxiv.org/abs/2311.01536}{{\tt 2311.01536}}.

\bibitem{Hull:1986kz}
C.~M. Hull, \emph{{Compactifications of the Heterotic Superstring}}, \href{http://dx.doi.org/10.1016/0370-2693(86)91393-6}{\emph{Phys. Lett. B} {\bf 178} (1986) 357--364}.

\bibitem{Strominger:1986uh}
A.~Strominger, \emph{{Superstrings with Torsion}}, \href{http://dx.doi.org/10.1016/0550-3213(86)90286-5}{\emph{Nucl. Phys. B} {\bf 274} (1986) 253}.

\bibitem{Melnikov:2010sa}
I.~V. Melnikov and M.~R. Plesser, \emph{{A (0,2) Mirror Map}}, \href{http://dx.doi.org/10.1007/JHEP02(2011)001}{\emph{JHEP} {\bf 02} (2011) 001}, [\href{http://arxiv.org/abs/1003.1303}{{\tt 1003.1303}}].

\bibitem{Melnikov:2012hk}
I.~Melnikov, S.~Sethi and E.~Sharpe, \emph{{Recent Developments in (0,2) Mirror Symmetry}}, \href{http://dx.doi.org/10.3842/SIGMA.2012.068}{\emph{SIGMA} {\bf 8} (2012) 068}, [\href{http://arxiv.org/abs/1209.1134}{{\tt 1209.1134}}].

\bibitem{Gu:2019byn}
W.~Gu, J.~Guo and E.~Sharpe, \emph{{A proposal for nonabelian $(0,2)$ mirrors}}, \href{http://dx.doi.org/10.4310/ATMP.2021.v25.n6.a4}{\emph{Adv. Theor. Math. Phys.} {\bf 25} (2021) 1549--1596}, [\href{http://arxiv.org/abs/1908.06036}{{\tt 1908.06036}}].

\bibitem{Anderson:2010mh}
L.~B. Anderson, J.~Gray, A.~Lukas and B.~Ovrut, \emph{{Stabilizing the Complex Structure in Heterotic Calabi-Yau Vacua}}, \href{http://dx.doi.org/10.1007/JHEP02(2011)088}{\emph{JHEP} {\bf 02} (2011) 088}, [\href{http://arxiv.org/abs/1010.0255}{{\tt 1010.0255}}].

\bibitem{Anderson:2011ty}
L.~B. Anderson, J.~Gray, A.~Lukas and B.~Ovrut, \emph{{The Atiyah Class and Complex Structure Stabilization in Heterotic Calabi-Yau Compactifications}}, \href{http://dx.doi.org/10.1007/JHEP10(2011)032}{\emph{JHEP} {\bf 10} (2011) 032}, [\href{http://arxiv.org/abs/1107.5076}{{\tt 1107.5076}}].

\bibitem{Anderson:2014xha}
L.~B. Anderson, J.~Gray and E.~Sharpe, \emph{{Algebroids, Heterotic Moduli Spaces and the Strominger System}}, \href{http://dx.doi.org/10.1007/JHEP07(2014)037}{\emph{JHEP} {\bf 07} (2014) 037}, [\href{http://arxiv.org/abs/1402.1532}{{\tt 1402.1532}}].

\bibitem{Garcia-Fernandez:2015hja}
M.~Garcia-Fernandez, R.~Rubio and C.~Tipler, \emph{{Infinitesimal moduli for the Strominger system and Killing spinors in generalized geometry}}, \href{http://dx.doi.org/10.1007/s00208-016-1463-5}{\emph{Math. Ann.} {\bf 369} (2017) 539--595}, [\href{http://arxiv.org/abs/1503.07562}{{\tt 1503.07562}}].

\bibitem{Ashmore:2018ybe}
A.~Ashmore, X.~De~La~Ossa, R.~Minasian, C.~Strickland-Constable and E.~E. Svanes, \emph{{Finite deformations from a heterotic superpotential: holomorphic Chern-Simons and an $L_\infty$ algebra}}, \href{http://dx.doi.org/10.1007/JHEP10(2018)179}{\emph{JHEP} {\bf 10} (2018) 179}, [\href{http://arxiv.org/abs/1806.08367}{{\tt 1806.08367}}].

\bibitem{Ashmore:2019rkx}
A.~Ashmore, C.~Strickland-Constable, D.~Tennyson and D.~Waldram, \emph{{Heterotic backgrounds via generalised geometry: moment maps and moduli}}, \href{http://dx.doi.org/10.1007/JHEP11(2020)071}{\emph{JHEP} {\bf 11} (2020) 071}, [\href{http://arxiv.org/abs/1912.09981}{{\tt 1912.09981}}].

\bibitem{McOrist:2021dnd}
J.~McOrist and E.~E. Svanes, \emph{{Heterotic quantum cohomology}}, \href{http://dx.doi.org/10.1007/JHEP11(2022)096}{\emph{JHEP} {\bf 11} (2022) 096}, [\href{http://arxiv.org/abs/2110.06549}{{\tt 2110.06549}}].

\bibitem{Bershadsky:1993cx}
M.~Bershadsky, S.~Cecotti, H.~Ooguri and C.~Vafa, \emph{{Kodaira-Spencer theory of gravity and exact results for quantum string amplitudes}}, \href{http://dx.doi.org/10.1007/BF02099774}{\emph{Commun. Math. Phys.} {\bf 165} (1994) 311--428}, [\href{http://arxiv.org/abs/hep-th/9309140}{{\tt hep-th/9309140}}].

\bibitem{donaldson1998gauge}
S.~K. Donaldson and R.~P. Thomas, \emph{{Gauge Theory in Higher Dimensions}},  in \emph{{The Geometric Universe: Science, Geometry, and the Work of Roger Penrose}}.
\newblock Oxford University Press, 04, 1998.
\newblock \href{http://dx.doi.org/10.1093/oso/9780198500599.003.0003}{DOI}.

\bibitem{thomas1997gauge}
R.~P. Thomas, \emph{Gauge theory on Calabi-Yau manifolds}.
\newblock PhD thesis, University of Oxford, 1997.

\bibitem{Garcia-Fernandez:2018emx}
M.~Garcia-Fernandez, R.~Rubio, C.~Shahbazi and C.~Tipler, \emph{{Canonical metrics on holomorphic Courant algebroids}}, \href{http://dx.doi.org/10.1112/plms.12468}{\emph{Proc. Lond. Math. Soc.} {\bf 125} (2022) 700--758}, [\href{http://arxiv.org/abs/1803.01873}{{\tt 1803.01873}}].

\bibitem{Garcia-Fernandez:2020awc}
M.~Garcia-Fernandez, R.~Rubio and C.~Tipler, \emph{{Gauge theory for string algebroids}},  \href{http://arxiv.org/abs/2004.11399}{{\tt 2004.11399}}.

\bibitem{Tellez-Dominguez:2023wwr}
R.~Tellez-Dominguez, \emph{{Chern correspondence for higher principal bundles}},  \href{http://arxiv.org/abs/2310.12738}{{\tt 2310.12738}}.

\bibitem{Streets:2024rfo}
J.~Streets, C.~Strickland-Constable and F.~Valach, \emph{{Ricci flow on Courant algebroids}},  \href{http://arxiv.org/abs/2402.11069}{{\tt 2402.11069}}.

\bibitem{Silva:2024fvl}
A.~A.~d. Silva, M.~Garcia-Fernandez, J.~D. Lotay and H.~N.~S. Earp, \emph{{Coupled $\operatorname{G}_2$-instantons}},  \href{http://arxiv.org/abs/2404.12937}{{\tt 2404.12937}}.

\bibitem{Lee:2019xtm}
S.-J. Lee, W.~Lerche and T.~Weigand, \emph{{Emergent strings, duality and weak coupling limits for two-form fields}}, \href{http://dx.doi.org/10.1007/JHEP02(2022)096}{\emph{JHEP} {\bf 02} (2022) 096}, [\href{http://arxiv.org/abs/1904.06344}{{\tt 1904.06344}}].

\bibitem{LopesCardoso:2003dvb}
G.~Lopes~Cardoso, G.~Curio, G.~Dall'Agata and D.~Lust, \emph{{BPS action and superpotential for heterotic string compactifications with fluxes}}, \href{http://dx.doi.org/10.1088/1126-6708/2003/10/004}{\emph{JHEP} {\bf 10} (2003) 004}, [\href{http://arxiv.org/abs/hep-th/0306088}{{\tt hep-th/0306088}}].

\bibitem{Becker:2003gq}
K.~Becker, M.~Becker, K.~Dasgupta and S.~Prokushkin, \emph{{Properties of heterotic vacua from superpotentials}}, \href{http://dx.doi.org/10.1016/S0550-3213(03)00495-4}{\emph{Nucl. Phys. B} {\bf 666} (2003) 144--174}, [\href{http://arxiv.org/abs/hep-th/0304001}{{\tt hep-th/0304001}}].

\bibitem{Gurrieri:2004dt}
S.~Gurrieri, A.~Lukas and A.~Micu, \emph{{Heterotic on half-flat}}, \href{http://dx.doi.org/10.1103/PhysRevD.70.126009}{\emph{Phys. Rev. D} {\bf 70} (2004) 126009}, [\href{http://arxiv.org/abs/hep-th/0408121}{{\tt hep-th/0408121}}].

\bibitem{Ashmore:2023vji}
A.~Ashmore, J.~J.~M. Ibarra, D.~D. McNutt, C.~Strickland-Constable, E.~E. Svanes, D.~Tennyson et~al., \emph{{A heterotic Kodaira-Spencer theory at one-loop}}, \href{http://dx.doi.org/10.1007/JHEP10(2023)130}{\emph{JHEP} {\bf 10} (2023) 130}, [\href{http://arxiv.org/abs/2306.10106}{{\tt 2306.10106}}].

\bibitem{schweitzer2007autour}
M.~Schweitzer, \emph{Autour de la cohomologie de bott-chern}, {\emph{arXiv preprint arXiv:0709.3528} (2007) }.

\bibitem{Dine:1985rz}
M.~Dine, R.~Rohm, N.~Seiberg and E.~Witten, \emph{{Gluino Condensation in Superstring Models}}, \href{http://dx.doi.org/10.1016/0370-2693(85)91354-1}{\emph{Phys. Lett. B} {\bf 156} (1985) 55--60}.

\bibitem{derendinger1985low}
J.-P. Derendinger, L.~E. Ibanez and H.~P. Nilles, \emph{On the low energy d= 4, n= 1 supergravity theory extracted from the d= 10, n= 1 superstring}, {\emph{Physics Letters B} {\bf 155} (1985) 65--70}.

\bibitem{LopesCardoso:2003sp}
G.~Lopes~Cardoso, G.~Curio, G.~Dall'Agata and D.~Lust, \emph{{Heterotic string theory on nonKahler manifolds with H flux and gaugino condensate}}, \href{http://dx.doi.org/10.1002/prop.200310134}{\emph{Fortsch. Phys.} {\bf 52} (2004) 483--488}, [\href{http://arxiv.org/abs/hep-th/0310021}{{\tt hep-th/0310021}}].

\bibitem{Manousselis:2005xa}
P.~Manousselis, N.~Prezas and G.~Zoupanos, \emph{{Supersymmetric compactifications of heterotic strings with fluxes and condensates}}, \href{http://dx.doi.org/10.1016/j.nuclphysb.2006.01.008}{\emph{Nucl. Phys. B} {\bf 739} (2006) 85--105}, [\href{http://arxiv.org/abs/hep-th/0511122}{{\tt hep-th/0511122}}].

\bibitem{Chatzistavrakidis:2012qb}
A.~Chatzistavrakidis, O.~Lechtenfeld and A.~D. Popov, \emph{{Nearly K\"ahler heterotic compactifications with fermion condensates}}, \href{http://dx.doi.org/10.1007/JHEP04(2012)114}{\emph{JHEP} {\bf 04} (2012) 114}, [\href{http://arxiv.org/abs/1202.1278}{{\tt 1202.1278}}].

\bibitem{Minasian:2017eur}
R.~Minasian, M.~Petrini and E.~E. Svanes, \emph{{On Heterotic Vacua with Fermionic Expectation Values}}, \href{http://dx.doi.org/10.1002/prop.201700010}{\emph{Fortsch. Phys.} {\bf 65} (2017) 1700010}, [\href{http://arxiv.org/abs/1702.01156}{{\tt 1702.01156}}].

\bibitem{Candelas:1990pi}
P.~Candelas and X.~de~la Ossa, \emph{{Moduli Space of {Calabi-Yau} Manifolds}}, \href{http://dx.doi.org/10.1016/0550-3213(91)90122-E}{\emph{Nucl. Phys. B} {\bf 355} (1991) 455--481}.

\bibitem{Strominger:1990pd}
A.~Strominger, \emph{{SPECIAL GEOMETRY}}, \href{http://dx.doi.org/10.1007/BF02096559}{\emph{Commun. Math. Phys.} {\bf 133} (1990) 163--180}.

\bibitem{Candelas:1990rm}
P.~Candelas, X.~C. De~La~Ossa, P.~S. Green and L.~Parkes, \emph{{A Pair of Calabi-Yau manifolds as an exactly soluble superconformal theory}}, \href{http://dx.doi.org/10.1016/0550-3213(91)90292-6}{\emph{Nucl. Phys. B} {\bf 359} (1991) 21--74}.

\bibitem{Kollar:2012pv}
J.~Kollar, \emph{{Deformations of elliptic Calabi-Yau manifolds}},  \href{http://arxiv.org/abs/1206.5721}{{\tt 1206.5721}}.

\bibitem{Oguiso1993ONAF}
K.~Oguiso, \emph{On algebraic fiber space structures on a calabi-yau 3-fold}, {\emph{International Journal of Mathematics} {\bf 04} (1993) 439--465}.

\bibitem{Wilson1994}
P.~Wilson, \emph{The existence of elliptic fibre space structures on calabi-yau threefolds.}, {\emph{Mathematische Annalen} {\bf 300} (1994) 693--704}.

\bibitem{Grimm:2019bey}
T.~W. Grimm, F.~Ruehle and D.~van~de Heisteeg, \emph{{Classifying Calabi\textendash{}Yau Threefolds Using Infinite Distance Limits}}, \href{http://dx.doi.org/10.1007/s00220-021-03972-9}{\emph{Commun. Math. Phys.} {\bf 382} (2021) 239--275}, [\href{http://arxiv.org/abs/1910.02963}{{\tt 1910.02963}}].

\bibitem{Grimm:2018cpv}
T.~W. Grimm, C.~Li and E.~Palti, \emph{{Infinite Distance Networks in Field Space and Charge Orbits}}, \href{http://dx.doi.org/10.1007/JHEP03(2019)016}{\emph{JHEP} {\bf 03} (2019) 016}, [\href{http://arxiv.org/abs/1811.02571}{{\tt 1811.02571}}].

\bibitem{Bonezzi:2024dlv}
R.~Bonezzi, F.~Diaz-Jaramillo and O.~Hohm, \emph{{Double Copy of 3D Chern-Simons Theory and 6D Kodaira-Spencer Gravity}},  \href{http://arxiv.org/abs/2404.16830}{{\tt 2404.16830}}.

\bibitem{Tseng:2009gr}
L.-S. Tseng and S.-T. Yau, \emph{{Cohomology and Hodge Theory on Symplectic Manifolds. I.}}, {\emph{J. Diff. Geom.} {\bf 91} (2012) 383--416}, [\href{http://arxiv.org/abs/0909.5418}{{\tt 0909.5418}}].

\bibitem{Tseng:2010kt}
L.-S. Tseng and S.-T. Yau, \emph{{Cohomology and Hodge Theory on Symplectic Manifolds. II}}, {\emph{J. Diff. Geom.} {\bf 91} (2012) 417--443}, [\href{http://arxiv.org/abs/1011.1250}{{\tt 1011.1250}}].

\bibitem{Tsai:2014ela}
C.-J. Tsai, L.-S. Tseng and S.-T. Yau, \emph{{Cohomology and Hodge Theory on Symplectic Manifolds: III}}, {\emph{J. Diff. Geom.} {\bf 103} (2016) 83--143}, [\href{http://arxiv.org/abs/1402.0427}{{\tt 1402.0427}}].

\bibitem{Tseng:2011gv}
L.-S. Tseng and S.-T. Yau, \emph{{Generalized Cohomologies and Supersymmetry}}, \href{http://dx.doi.org/10.1007/s00220-014-1895-2}{\emph{Commun. Math. Phys.} {\bf 326} (2014) 875--885}, [\href{http://arxiv.org/abs/1111.6968}{{\tt 1111.6968}}].

\bibitem{Lau:2014fia}
S.-C. Lau, L.-S. Tseng and S.-T. Yau, \emph{{Non-K\"ahler SYZ Mirror Symmetry}}, \href{http://dx.doi.org/10.1007/s00220-015-2454-1}{\emph{Commun. Math. Phys.} {\bf 340} (2015) 145--170}, [\href{http://arxiv.org/abs/1409.2765}{{\tt 1409.2765}}].

\bibitem{Marchesano:2014iea}
F.~Marchesano, D.~Regalado and G.~Zoccarato, \emph{{On D-brane moduli stabilisation}}, \href{http://dx.doi.org/10.1007/JHEP11(2014)097}{\emph{JHEP} {\bf 11} (2014) 097}, [\href{http://arxiv.org/abs/1410.0209}{{\tt 1410.0209}}].

\bibitem{Gray:2018kss}
J.~Gray and H.~Parsian, \emph{{Moduli identification methods in Type II compactifications}}, \href{http://dx.doi.org/10.1007/JHEP07(2018)158}{\emph{JHEP} {\bf 07} (2018) 158}, [\href{http://arxiv.org/abs/1803.08176}{{\tt 1803.08176}}].

\bibitem{bedulli2023syz}
L.~Bedulli and A.~Vannini, \emph{Syz mirror symmetry of solvmanifolds}, {\emph{arXiv e-prints} (2023) arXiv--2311}.

\bibitem{Kupka:2024rvl}
J.~Kupka, C.~Strickland-Constable, E.~E. Svanes, D.~Tennyson and F.~Valach, \emph{{BPS complexes and Chern--Simons theories from $G$-structures in gauge theory and gravity}},  \href{http://arxiv.org/abs/2406.03550}{{\tt 2406.03550}}.

\bibitem{Strominger:1996it}
A.~Strominger, S.-T. Yau and E.~Zaslow, \emph{{Mirror symmetry is T duality}}, \href{http://dx.doi.org/10.1016/0550-3213(96)00434-8}{\emph{Nucl. Phys. B} {\bf 479} (1996) 243--259}, [\href{http://arxiv.org/abs/hep-th/9606040}{{\tt hep-th/9606040}}].

\bibitem{Acharya:2001gy}
B.~S. Acharya and E.~Witten, \emph{{Chiral fermions from manifolds of G(2) holonomy}},  \href{http://arxiv.org/abs/hep-th/0109152}{{\tt hep-th/0109152}}.

\bibitem{Pantev:2009de}
T.~Pantev and M.~Wijnholt, \emph{{Hitchin's Equations and M-Theory Phenomenology}}, \href{http://dx.doi.org/10.1016/j.geomphys.2011.02.014}{\emph{J. Geom. Phys.} {\bf 61} (2011) 1223--1247}, [\href{http://arxiv.org/abs/0905.1968}{{\tt 0905.1968}}].

\bibitem{Braun:2018vhk}
A.~P. Braun, S.~Cizel, M.~H\"ubner and S.~Sch\"afer-Nameki, \emph{{Higgs bundles for M-theory on $G_{2}$-manifolds}}, \href{http://dx.doi.org/10.1007/JHEP03(2019)199}{\emph{JHEP} {\bf 03} (2019) 199}, [\href{http://arxiv.org/abs/1812.06072}{{\tt 1812.06072}}].

\bibitem{Barbosa:2019bgh}
R.~Barbosa, M.~Cveti\v{c}, J.~J. Heckman, C.~Lawrie, E.~Torres and G.~Zoccarato, \emph{{T-branes and $G_2$ backgrounds}}, \href{http://dx.doi.org/10.1103/PhysRevD.101.026015}{\emph{Phys. Rev. D} {\bf 101} (2020) 026015}, [\href{http://arxiv.org/abs/1906.02212}{{\tt 1906.02212}}].

\bibitem{Barbosa:2019hts}
R.~Barbosa, \emph{{Harmonic Higgs Bundles and Coassociative ALE Fibrations}},  \href{http://arxiv.org/abs/1910.10742}{{\tt 1910.10742}}.

\bibitem{Cvetic:2020piw}
M.~Cveti\v{c}, J.~J. Heckman, T.~B. Rochais, E.~Torres and G.~Zoccarato, \emph{{Geometric unification of Higgs bundle vacua}}, \href{http://dx.doi.org/10.1103/PhysRevD.102.106012}{\emph{Phys. Rev. D} {\bf 102} (2020) 106012}, [\href{http://arxiv.org/abs/2003.13682}{{\tt 2003.13682}}].

\bibitem{Acharya:2020xgn}
B.~S. Acharya, A.~Kinsella and E.~E. Svanes, \emph{{$T^{3}$-invariant heterotic Hull-Strominger solutions}}, \href{http://dx.doi.org/10.1007/JHEP01(2021)197}{\emph{JHEP} {\bf 01} (2021) 197}, [\href{http://arxiv.org/abs/2010.07438}{{\tt 2010.07438}}].

\bibitem{Hubner:2020yde}
M.~Hubner, \emph{{Local G$_{2}$-manifolds, Higgs bundles and a colored quantum mechanics}}, \href{http://dx.doi.org/10.1007/JHEP05(2021)002}{\emph{JHEP} {\bf 05} (2021) 002}, [\href{http://arxiv.org/abs/2009.07136}{{\tt 2009.07136}}].

\bibitem{Acharya:2021rvh}
B.~S. Acharya, A.~Kinsella and D.~R. Morrison, \emph{{Non-perturbative heterotic duals of M-theory on G$_{2}$ orbifolds}}, \href{http://dx.doi.org/10.1007/JHEP11(2021)065}{\emph{JHEP} {\bf 11} (2021) 065}, [\href{http://arxiv.org/abs/2106.03886}{{\tt 2106.03886}}].

\bibitem{Acharya:2023xlx}
B.~S. Acharya and D.~A. Baldwin, \emph{{Coulomb and Higgs phases of G$_{2}$-manifolds}}, \href{http://dx.doi.org/10.1007/JHEP01(2024)147}{\emph{JHEP} {\bf 01} (2024) 147}, [\href{http://arxiv.org/abs/2309.12869}{{\tt 2309.12869}}].

\bibitem{Acharya:2023syl}
B.~S. Acharya and D.~A. Baldwin, \emph{{Ricci Flat Metrics, Flat Connections and $G_{2}$-Manifolds}},  \href{http://arxiv.org/abs/2312.12311}{{\tt 2312.12311}}.

\bibitem{Alvarez-Garcia:2021pxo}
R.~\'Alvarez-Garc\'\i{}a, D.~Kl\"awer and T.~Weigand, \emph{{Membrane limits in quantum gravity}}, \href{http://dx.doi.org/10.1103/PhysRevD.105.066024}{\emph{Phys. Rev. D} {\bf 105} (2022) 066024}, [\href{http://arxiv.org/abs/2112.09136}{{\tt 2112.09136}}].

\bibitem{Beasley:2003fx}
C.~Beasley and E.~Witten, \emph{{Residues and world sheet instantons}}, \href{http://dx.doi.org/10.1088/1126-6708/2003/10/065}{\emph{JHEP} {\bf 10} (2003) 065}, [\href{http://arxiv.org/abs/hep-th/0304115}{{\tt hep-th/0304115}}].

\bibitem{Buchbinder:2019hyb}
E.~I. Buchbinder, A.~Lukas, B.~A. Ovrut and F.~Ruehle, \emph{{Heterotic Instantons for Monad and Extension Bundles}}, \href{http://dx.doi.org/10.1007/JHEP02(2020)081}{\emph{JHEP} {\bf 02} (2020) 081}, [\href{http://arxiv.org/abs/1912.07222}{{\tt 1912.07222}}].

\bibitem{Buchbinder:2019eal}
E.~I. Buchbinder, A.~Lukas, B.~A. Ovrut and F.~Ruehle, \emph{{Instantons and Hilbert Functions}}, \href{http://dx.doi.org/10.1103/PhysRevD.102.026019}{\emph{Phys. Rev. D} {\bf 102} (2020) 026019}, [\href{http://arxiv.org/abs/1912.08358}{{\tt 1912.08358}}].

\bibitem{Dine:1986zy}
M.~Dine, N.~Seiberg, X.~G. Wen and E.~Witten, \emph{{Nonperturbative Effects on the String World Sheet}}, \href{http://dx.doi.org/10.1016/0550-3213(86)90418-9}{\emph{Nucl. Phys. B} {\bf 278} (1986) 769--789}.

\bibitem{Dine:1987bq}
M.~Dine, N.~Seiberg, X.~G. Wen and E.~Witten, \emph{{Nonperturbative Effects on the String World Sheet. 2.}}, \href{http://dx.doi.org/10.1016/0550-3213(87)90383-X}{\emph{Nucl. Phys. B} {\bf 289} (1987) 319--363}.

\end{thebibliography}\endgroup

\end{document}